\documentclass[superscriptaddress,groupedaddress,nofootnoteinbib,11pt]{article}
\pdfoutput=1 

\usepackage{jcappub}
\usepackage{bm}
\usepackage{verbatim}

\makeatletter
\gdef\@fpheader{}
\g@addto@macro\bfseries{\boldmath}
\makeatother

\usepackage{epsf}
\usepackage{graphicx,epsfig}
\usepackage{amsfonts}
\usepackage{amssymb}
\usepackage{xcolor}
\usepackage{paralist} 

\usepackage{ulem}

\definecolor{mine}{rgb}{0.2,0.1,0.7}
\definecolor{bb}{rgb}{0.3, 0.5, 1}
\definecolor{bg}{rgb}{0.1, 0.1, 0.5}

\def\half{\frac12}
\def\M{M_{{}_{\mathrm{Pl}}}}

\def\({\left(}
\def\){\right)}
\def\ie{{\it i.e. }}

\def\R{R_{\rm fs}}

\def\kc{k_\mathrm{c}}

\def\Ndes{\Delta N_{{\rm GD}}}
\def\DeltaN{\Delta N_{\ell,\mathrm{c}}}

\newcommand{\bea}{\begin{eqnarray}}
\newcommand{\eea}{\end{eqnarray}}
\newcommand\be{\begin{equation}}
\newcommand\ee{\end{equation}}
\newcommand\beq{\begin{equation}}
\newcommand\eeq{\end{equation}}

\def\ba{\begin{eqnarray}}
\def\ea{\end{eqnarray}}

\newcommand{\Eq}[1]{Eq.~(\ref{#1})}
\newcommand{\Eqs}[1]{Eqs.~(\ref{#1})}
\newcommand{\Fig}[1]{Fig.~{\ref{#1}}}
\newcommand{\Figs}[1]{Figs.~{\ref{#1}}}
\newcommand{\Ref}[1]{Ref.~{\cite{#1}}}
\newcommand{\Refs}[1]{Refs.~{\cite{#1}}}
\newcommand{\Sec}[1]{Sec.~\ref{#1}}

\renewcommand{\imath}{\mathrm{i}}

\newcommand{\refeq}[1]{(\ref{#1})}

\def\Tdot#1{{{#1}^{\hbox{.}}}}

\begin{document}

\title{On backreaction effects in geometrical destabilisation of inflation}

\author[1]{Oskar Grocholski,}
\author[1]{Marcin Kalinowski,}
\author[1]{Maciej Kolanowski,}
\author[2,3]{S\'ebastien Renaux-Petel,} 
\author[1]{Krzysztof Turzy\'nski,}
\author[4,2]{Vincent Vennin.}
\affiliation[1]{Institute of Theoretical Physics, Faculty of Physics, University of Warsaw, Pasteura 5, \mbox{02-093 Warsaw}, Poland}
\affiliation[2]{Institut d'Astrophysique de Paris, UMR-7095 du CNRS, Universit\'e Pierre et Marie Curie, 98~bis~bd~Arago, 75014 Paris, France.}
\affiliation[3]{Sorbonne Universit\'es, Institut Lagrange de Paris, 98 bis bd Arago, 75014 Paris, France}
\affiliation[4]{Laboratoire Astroparticule et Cosmologie, Universit\'e Denis Diderot Paris 7, 75013 Paris,
France}

\emailAdd{oskar.grocholski@gmail.com, mj.kalinowski@student.uw.edu.pl, maciej.pawel.kolanowski@gmail.com, renaux@iap.fr, krzysztof-jan.turzynski@fuw.edu.pl, vincent.vennin@apc.in2p3.fr}

\vskip 8pt

\date{\today}


\abstract{We study the geometrical instability arising in multi-field models of inflation with negatively-curved field space. We analyse how the homogeneous background evolves in presence of geometrical destabilisation, and show that, in simple models, a kinematical backreaction effect takes place that shuts off the instability. We also follow the evolution of the unstable scalar fluctuations and show that, in most situations, they must remain in the perturbative regime in order to satisfy observational constraints. We conclude that, in the simplest models of geometrical destabilisation, inflation does not end prematurely, but rather proceeds along a modified, sidetracked, field-space trajectory.}

\maketitle

\section{Introduction}

Cosmological inflation~\cite{Starobinsky:1980te, Sato:1980yn, Guth:1980zm, Linde:1981mu, Albrecht:1982wi, Linde:1983gd}
 solves many problems in big-bang cosmology, including generation of primordial density perturbations out of
 amplified quantum fluctuations of the gravitational and matter fields~\cite{Starobinsky:1979ty, Mukhanov:1981xt, Hawking:1982cz,  Starobinsky:1982ee, Guth:1982ec, Bardeen:1983qw}.
Therefore, inflation has become a natural ingredient of the standard cosmological model. 
At present, the simplest inflationary models, with a single scalar inflaton which has a canonical kinetic term and is minimally coupled to gravity,
is consistent with observational data~\cite{Ade:2015lrj,Ade:2015ava, Martin:2013nzq, Martin:2013tda}. 
In most models, the inflaton field evolves in a flat potential in the slow-roll regime and when the potential becomes too steep inflation ends. 

Recently, it has been proposed that the field-space curvature of inflationary models with multiple fields can dominate forces originating from the potential gradient and destabilise inflationary trajectories, a phenomenon dubbed geometrical destabilisation \cite{RT}. It has however been unclear whether this instability leads to a premature end of inflation \cite{GD1} or to abandoning the effective
single-field regime \cite{GD2,Cicoli:2018ccr}. Both possibilities yield predictions for the inflationary observables, such as the scalar spectral index and the tensor-to-scalar ratio, markedly different than those
in single-field models; however the underlying physics is very different. It is therefore important to understand better the dynamics of geometrical destabilisation by means of both analytical techniques and 
numerical simulations.

In this work, we make a first step in this direction, by studying analytically the evolution of the homogeneous background and the growth of the unstable perturbations in the linear regime. Our work is
organised as follows.  In \Sec{sec:ess}, we briefly review multi-field inflationary models with non-canonical kinetic terms, and geometrical destabilisation which may occur in these models. 
In \Sec{sec:kin_all}, we study how the homogeneous background evolves in the presence of geometrical destabilisation and show that the resulting kinematical effects lead to shutting off the
instability. In \Sec{sec:linana}, we follow the evolution of the unstable scalar field perturbations and show that 
they are unlikely to end inflation.
We conclude in \Sec{sec:sum}.

\section{Essentials of geometrical destabilisation}
\label{sec:ess}

\subsection{General mechanism} 
Geometrical destabilisation relies on the fact that when inflation is embedded in high-energy physics, it relies on actions for scalar fields that often contain kinetic terms of the type ${\cal L}_{\rm kin}=-\frac12 G_{IJ}(\phi^K) g^{\mu \nu} \partial_{\mu} \phi^I \partial_{\nu} \phi^J$, where the manifold described by the field-space metric $G_{IJ}$ is curved, \textit{i.e.} has a non-trivial Riemann curvature tensor. In the following, we will concentrate on non-linear sigma models involving an arbitrary number $N$ of scalar fields minimally coupled to Einstein gravity, and whose action reads
\be
S=\int {\rm d}^4 x \sqrt{-g} \left[-\frac12 G_{IJ}\left(\phi^K\right) \partial_{\mu} \phi^I \partial^{\mu} \phi^J-V\left(\phi^K\right) \right]\,.
\label{action}
\ee
From a top-down perspective, such as in inflationary model building in supergravity or in string theory, actions of the type \refeq{action} with a curved field space are ubiquitous. For instance, in ${\cal N}=1$ supergravity in four dimensions, the bosonic fields are the metric $g_{\mu \nu}$, gauge potentials $A_{\mu}^a$, and complex scalar fields $\phi^i$, whose low-energy Lagrangian reads, in the absence of gauge interactions:
\be
{\cal L}=-K_{i \bar j} \partial^\mu \phi^i \partial_\mu \bar \phi^j - V_{\rm F}.
\label{sugra}
\ee
Here, $K_{i \bar j} \equiv \frac{\partial^2 K}{\partial \phi^i \partial \bar \phi^j}$ is the so-called K\"ahler metric derived from the K\"ahler potential $K(\phi^i, \bar \phi^i)$, which is a real analytic function of the fields. The ${\rm F}$-term potential $V_{{\rm F}}$ is (see \textit{e.g.} \cite{BM})
\be
V_{{\rm F}}=e^{K/\M^2} \left(  K^{i \bar j} D_i W D_{\bar{j}} W -\frac{3}{\M^2} |W|^2 \right)\,,
\ee
where $K^{i \bar j}$ is the inverse K\"ahler metric and $D_i W \equiv \partial_i W+\frac{1}{\M^2} (\partial_i K) W$, where the superpotential $W(\phi^i)$ is a holomorphic function of the fields $\phi^i$. Upon expressing the Lagrangian \refeq{sugra} in terms of real scalar fields, it is indeed of the type \refeq{action}, though with a specific structure dictated by the superpotential and the K\"ahler potential of the theory. Note that in this context, the description of inflation using multiple fields is a built-in feature, as even the simplest models involve one \textit{complex} scalar field, \textit{i.e.} two real scalar fields. We also stress that the fact that the K\"ahler metric generically describes a curved internal space is dictated by the theoretical structure of these theories.
For instance, in the case of a string compactification with ${\cal N}=1$ supersymmetry, $K$ and $W$ are related to geometric properties of the compactification.

From a bottom-up effective field theory point of view, non-trivial field-space manifolds also naturally emerge. In an effective field theory with cutoff scale $\Lambda$, we parametrise our ignorance about the UV physics, by making assumptions about the symmetries of the UV theory, and by writing down the most general effective Lagrangian consistent with these symmetries:
\be
{\cal L}_{{\rm eff}}\left(\phi^I\right)={\cal L}_{\ell}\left(\phi^I\right)+\sum_i c_i \frac{{\cal O}_i\left(\phi^I,\partial \phi^I, \ldots\right)}{\Lambda^{\delta_i-4}}\,.
\label{EFT}
\ee
Here, ${\cal L}_{\ell}(\phi^I)$ is the renormalisable Lagrangian of the light (compared to the cutoff $\Lambda$) degrees of freedom $\phi^I$, and the sum runs over all higher-order operators ${\cal O}_i$ of dimension $\delta_i$ allowed by the symmetries, with dimensionless Wilson coefficients $c_i$. 
The operators ${\cal O}_i$'s are constructed not only from the fields $\phi^I$ but also from their (space-time) derivatives, which may in general 
lead to a non-trivial field space metric.

The dynamics of non-linear sigma models of inflation described by the Lagrangian \refeq{action} have been extensively studied in the past two decades 
(see, e.g., \cite{Sasaki:1995aw,Mukhanov:1997fw,GrootNibbelink:2001qt}). On a spatially flat Friedmann-Lema\^itre-Robertson-Walker universe, with metric 
\begin{equation}
\mathrm{d}s^2=-\mathrm{d}t^2+a^2(t)\mathrm{d}{\vec x}^2\,,
\end{equation}
where $t$ is cosmic time and $a(t)$ denotes the scale factor, and with homogeneous scalar fields $ \phi^I$, the equations of motion take the form: 
\begin{eqnarray}
 3H^2 \M^2&=&\frac12 \dot \sigma^2+V\,,  \\
 \dot{H} \M^2&=&-\frac12\dot \sigma^2\,, \label{Hdot} \\
 \label{eq:scfi_eom}
{\cal D}_t \dot \phi^I  +3H  \dot \phi^I+G^{IJ} V_{,J}&=&0\, .
\end{eqnarray}
In these expressions, dots denote derivatives with respect to $t$, $H \equiv \dot a/a$ is the Hubble parameter, $\frac{1}{2}\dot \sigma^2 \equiv \frac{1}{2}G_{IJ} \dot \phi^I \dot \phi^J$ is the kinetic energy of the fields, and, hereafter, ${\cal D}_t A^I \equiv \dot{A^I} + \Gamma^I_{JK} \dot \phi^J A^K$ for a field-space vector $A^I$ (field-space indices are lowered and raised with the field-space metric and its inverse).

The behaviour of linear fluctuations about such a background is governed by the second-order action
 \begin{eqnarray}
S_{(2)}= \int  \mathrm{d}t\, \mathrm{d}^3x \,a^3\left(G_{IJ}\mathcal{D}_tQ^I\mathcal{D}_tQ^J-\frac{1}{a^2}G_{IJ}\partial_i Q^I \partial^i Q^J-M_{IJ}Q^IQ^J\right),
\label{S2}
\end{eqnarray}
where the $Q^I$'s are the field fluctuations in the spatially flat gauge and $M_{IJ}$ is a mass (squared) matrix. The equations of motion deduced from Eq.~\refeq{S2} take the simple form of generalised harmonic oscillators equations (in Fourier space)
\begin{eqnarray}
{\cal D}_t {\cal D}_t Q^I  +3H {\cal D}_t Q^I +\frac{k^2}{a^2} Q^I +M^I_{\,J} Q^J=0\,
\label{pert}
\end{eqnarray}
with Hubble friction, and whose crucial physical information lies in the mass matrix
\begin{eqnarray} 
\label{masssquared}
M^{I}_{\,J} &=& V^{I}_{; J} - {\mathcal{R}^{I}_{\,KLJ}\dot \phi^K \dot \phi^L} -\frac{1}{a^3 \M^2}\mathcal{D}_t\left(\frac{a^3}{H} \dot \phi^
I \dot \phi_J\right).
\end{eqnarray}
The first term, $ V_{;IJ} \equiv V_{,IJ}-\Gamma_{IJ}^K V_{,K}$, is the Hessian of the potential, and the last term comes from the backreaction of the metric fluctuations. The second term contains the Riemann tensor associated to the field-space metric, $\mathcal{R}^{I}_{\,KLJ}$, which may drive the mechanism of geometrical destabilisation.

To gain physical intuition, let us concentrate on the case of two-field inflationary models.
The analysis of this situation is simplified as, in a two-dimensional space, there is only one independent component of the Riemann tensor, \textit{i.e.} $R_{IJKL}=1/2\R \left(G_{IK}G_{JL}-G_{IL}G_{JK} \right)$, where $\R$ denotes the field-space Ricci scalar curvature. It is then convenient to project the equations of motion \refeq{pert} onto the adiabatic/entropic basis $(e_{\sigma}^I,e_{s}^I)$ \cite{Gordon:2000hv,GrootNibbelink:2001qt}, where  $e_{\sigma}^I \equiv \dot \phi^I /{\dot \sigma}$ is the unit vector pointing along the background trajectory in field space, and where $e_s^I$ is such that the basis $(e_{\sigma}^I,e_{s}^I)$ is orthonormal and right-handed for definiteness. The adiabatic perturbation $Q_{\sigma} \equiv e_{\sigma I} Q^I$ is directly proportional to the comoving curvature perturbation ${\cal R}=\frac{H}{{\dot \sigma}} Q_{\sigma}$, 
while the genuine multifield effects are embodied by the entropic fluctuation $Q_s$, perpendicular to the background trajectory. In this basis, the equations of motion take the form
\begin{eqnarray}
\label{Qsigma}
 \ddot{Q}_{\sigma}+3H
 \dot{Q}_{\sigma}+\left(\frac{k^2}{a^2}+m_{\sigma}^2\right)  Q_{\sigma} &=& \Tdot{\left(2 H \eta_\perp Q_{s}\right)}
-\left(\frac{\dot H}{H}
+\frac{V_{,\sigma}}{\dot \sigma }\right)  2 H \eta_\perp\, Q_{s}\,, \\
\ddot{Q}_s+3H \dot{Q}_{s}+\left(\frac{k^2}{a^2}+m_{s}^2\right)  Q_{s}&=&-2 \dot \sigma \eta_\perp \dot{{\cal R}}\,,
\label{Qs}
\end{eqnarray}
where
\begin{equation} \label{etaperpdefinition}
\eta_\perp  \equiv -\frac{V_{,s} }{H \dot \sigma}
\end{equation}
is the dimensionless parameter measuring the size of the coupling between the adiabatic mode and the entropic fluctuations, which is non-zero when the trajectory deviates from a geodesic in field space \cite{GrootNibbelink:2001qt}. Here $V_{,s}  \equiv e_{s}^I V_{,I}$ (and similarly for analogous quantities), the adiabatic mass (squared) is given by $m_{\sigma}^2/H^2=-\frac{3}{2}\epsilon_2+\ldots$
with the slow-roll parameters given by $\epsilon_1 \equiv -\frac{\dot H}{H^2}$, $\epsilon_2 =\frac{\dot \epsilon_1}{H \epsilon_1}$ and the ellipsis representing terms of higher order in the slow-roll parameters, and the entropic mass squared reads $m_s^2=V_{;ss}+1/2 \dot \sigma^2 \R-\left( H \eta_\perp \right)^2$.
In the super-Hubble limit, \textit{i.e.} when $k \ll a H$, the equation of motion for the curvature perturbations has a first order integral and simplifies considerably to 
\be
 \dot {\cal R} = 2 \eta_{\perp} \frac{H^2}{\dot \sigma} Q_s \label{Rdot}+{\cal O}\left(\frac{k^2}{a^2 H^2}\right),
\ee
where we recover that the curvature perturbation is conserved on super-Hubble scales in the single-field case --- as there is no entropic fluctuation then --- or when the background trajectory follows a field-space geodesic --- as $\eta_\perp=0$ then. Inserting Eq.~\refeq{Rdot} into Eq.~\refeq{Qs}, the super-Hubble equation for the entropic fluctuation simplifies as well to
\be
\ddot Q_s+3 H \dot Q_s +m^{2}_{s {\rm (eff)}} Q_s = 0 \label{eqQs}\,,
\ee
where we denote by $m^{2}_{s {\rm (eff)}}$ the effective entropic mass on super-Hubble scales 
\begin{equation}
\frac{m^{2}_{s {\rm (eff)}}}{H^2} \equiv \frac{V_{;ss}}{H^2} +3 \eta_\perp^2+ \epsilon_1  \, \R \M^2 \,.
\label{ms2}
\end{equation}
It contains three contributions: the Hessian, the bending and the geometrical terms, respectively \cite{Peterson:2011yt,Turzynski:2014tza}. 

From \Eqs{eqQs} and~\refeq{ms2}, 
the mechanism of geometrical destabilisation of inflation is readily identified: it corresponds to situations in which the geometrical contribution is negative and dominates the sum of the two other contributions, so that the entropic fluctuation is tachyonic, and the underlying background trajectory is unstable. As $\epsilon_1$ is a positive quantity, the geometrical destabilisation in two-field models can only arise in setups with a scalar curvature that is negative, which is related to the fact that this makes neighbouring geodesics diverge from one another. When the field-space curvature is positive, it renders the entropic fluctuations even more massive, and does not modify the standard picture. We hence consider only negatively-curved field spaces in the following.

\subsection{A minimal realisation}
\label{sec:minimal}

We now describe a simple realisation of the geometrical destabilisation of inflation. We start with a model of slow-roll inflation driven by a scalar field $\phi$ with canonical kinetic term and potential $V(\phi)$, with the Lagrangian ${\cal L}_{\phi}=-\frac12 (\partial \phi)^2-V(\phi)$.  We consider a typical situation in which an extra scalar field $\chi$ is thought to be stabilised at the bottom of its potential by a large mass, larger than the Hubble scale. This is described by the simple Lagrangian ${\cal L}_{\chi}=-\frac12 (\partial \chi)^2-\frac12 m^2 \chi^2$, where $m$ stands for the heavy mass, \textit{i.e.} $m^2 \gg H^2$. 
In this situation, if $\chi$ is initially displaced from its minimum at zero, it will rapidly rolls back towards $\chi=0$ like $1/a^{3/2}$, so that, after a transient regime, it can effectively be considered as stabilised. 
Let us now consider the impact of the dimension six operator ${\cal L}_{{\rm int}}\propto-(\partial \phi)^2 \chi^2/M^2$, where $M$ is a scale of new physics that lies well above the Hubble scale, $M \gg H$. Such an operator respects the (approximate) shift-symmetry of the inflaton and is therefore expected from an effective field theory point of view. Our total Lagrangian thus reads
\be
{\cal L}=-\frac12 (\partial \phi)^2  \left(1+2 \frac{\chi^2}{M^2} \right) -V(\phi)-\frac12 (\partial \chi)^2-\frac12 m^2 \chi^2\,.
\label{minimal}
\ee
We do not consider terms linear in $\chi$ so that the configuration $\chi=0$ we started with is indeed a solution of the equations of motion. We stress that terms that are higher-order in $\chi$, such that $\lambda \chi^4$ or $(\partial \phi)^2 \chi^4/M^4$, are generally expected, either in top-down realisations or from the effective field theoretic viewpoint. However, they are suppressed near the inflationary valley at $\chi=0$ and do not modify our discussion with respect to the triggering of the instability. As we shall see in the following sections, they however play an important role in understanding the fate of the instability once $\chi$ is kicked off from $\chi=0$. 
The dimension six operator generates a curved field space with metric $(1+2 \chi^2/M^2) ({\rm d} \phi)^2+ ({\rm d} \chi)^2$, whose Ricci scalar is negative and reads 
$R_{{\rm fs}}=-4/M^2 \times (1+2 \chi^2/M^2)^{-2}$.
Along the inflationary valley $\chi=0$, the entropic fluctuation $Q_s$, which then simply coincides with the fluctuation of $\chi$, thus acquires the effective mass \refeq{ms2}, \textit{i.e.}
\be
\label{eq:ms2_2}
m^{2}_{s {\rm (eff)}}=m^2-4 \epsilon_1 H^2 \left( \frac{\M}{M}\right)^2\,,
\ee
as we have here $V_{;ss}=m^2, \left(\R\right)_{| \chi=0}=-4/M^2$, and the inflationary trajectory along $\chi=0$ is a field-space geodesic, so that $\eta_\perp=0$.

Not all choices of potentials $V(\phi)$ and parameters leads to a destabilisation. For instance, if the kinetic energy density $\epsilon_1 H^2 \M^2=\half \dot \phi^2$ decreases during inflation, the effective mass~(\ref{eq:ms2_2}) increases and destabilisation does not occur. In the slow-roll regime, this arises for models with $2 V''V > {V'}^2$, like e.g.~$V(\phi)\propto\phi^p$ with $p>2$. Even when the kinetic energy density grows, if the hierarchy $ \M/M$ is not large enough, \textit{i.e.} for $M> 2 \M H_{{\rm end}}/m$, where $H_{{\rm end}}$ is the value of the Hubble scale when $\epsilon_1$ reaches one, destabilisation does not occur either.
However, for large classes of models, notably all the concave potentials preferred by the data, $\epsilon_1 H^2$ grows during inflation, and even a modest hierarchy between $M$ and $\M$ is sufficient to generate the instability, at the critical point such that 
\bea
\label{eq:epsilonc}
\epsilon_{1, {\rm c}} =  \frac{1}{4} \left(\frac{m}{H_{\rm c}}\right)^2 \left(\frac{M}{\M}\right)^2\, .
\eea
We consider such models in the following, and note for future use that the condition that $\epsilon_1 H^2$ grows during inflation can be simply restated as $\epsilon_2 > 2 \epsilon_1$.

The fact that the super-Hubble entropic mass becomes negative at a certain critical point simply means that the initial background trajectory  at $\chi=0$ is classically unstable once this point is reached.
If we denote by $\kc$ the Fourier mode that crosses the Hubble radius at the critical point, all the entropic fluctuations whose wavelengths are larger, \textit{i.e.} with $k \lesssim \kc$, experience the same exponential growth after the critical point (we shall discuss this issue quantitatively in \Sec{sec:linana}). The subsequent evolution depends on the backreaction of these fluctuations on the inflationary trajectory, a challenging and model-dependent problem that we address in the remainder of the paper.

Let us eventually remark that the action (\ref{minimal}) is a special case of a simplified, but well-motivated two-field action that has been used in the past
by many authors (see, {\it e.g.}\ \cite{mf1,mf2,mf3,mf4,Tolley:2009fg}) 
to study the effects of non-trivial field space curvature on the evolution of the perturbations: 
\be
{\cal L}=-\frac12 e^{2b(\chi)}(\partial \phi)^2   -V(\phi)-\frac12 (\partial \chi)^2-\frac12 m^2 \chi^2\,.
\label{minimal_two}
\ee
It will turn out useful to compare the predictions of the minimal action~(\ref{minimal}) with those of its slight generalisation (\ref{minimal_two}), to determine which effects are generic and which are peculiar to our minimal setup.

\section{Evolution of homogeneous fields}  
\label{sec:kin_all}

\subsection{Initial conditions for the spectator field}
\label{sec:ini}

The initial displacement of the spectator field $\chi$ is a crucial parameter for determining the duration of geometrical destabilisation. 
Classically, if $\chi$ is stabilised for a long period prior to destabilisation, its vacuum expectation value rolls down to 
tiny values.
However, as soon as it becomes light, quantum fluctuations source its large-scale component as they cross out the Hubble radius and provide the main contribution to its mean displacement. 
Let us then focus on the time interval during which the field is light (in the sense $m_{s,{\rm (eff)}}^2<H^2$) but still stabilised ($m_{s,{\rm (eff)}}^2>0$).  
In the slow-roll regime, the dynamics of $\chi$ coarse-grained on super-Hubble scales can be described with the Langevin equation~\cite{Starobinsky:1986fx}
\be
\frac{\mathrm{d}\chi}{\mathrm{d} N} + \frac{m_{s,{\rm (eff)}}^2}{3H^2}\chi = \frac{H}{2\pi}\xi\,,
\label{simple-stochastic}
\ee
where $\xi$ is a Gaussian white noise with unit variance, \ie such that $\langle \xi(N)\xi(N')\rangle = \delta(N-N')$, and $N=\ln a$ denotes the number of $e$-folds of expansion. One should be aware that this simple equation comes with its limitations. For instance, there are corrections to the noise amplitude, especially when $m_{s,{\rm (eff)}}^2$ approaches $H^2$. Also, $m_{s,{\rm (eff)}}^2$ varies rapidly compared to the scale factor as one approaches the critical time, so that one may expect non-negligible corrections to the slow-roll approximation, calling for a full phase space study. However, it is beyond the scope of this work to consider these aspects, and in the following, we simply use \Eq{simple-stochastic} in order to give a well-motivated estimate of the initial displacement of the spectator field $\chi$. Note eventually that here, $H$ is not a function of the stochastic field $\chi$, but rather a known function of $N$, so that the noise is not multiplicative and there is no It\^o versus Stratonovich ambiguity \cite{Pinol:2018euk}.
The solution of \Eq{simple-stochastic} reads
\be
\chi(N) = \int_{N_\ell}^N \mathrm{d} N_1 \frac{H(N_1)}{2\pi} \xi(N_1)\exp\left[-\int_{N_1}^{N}\mathrm{d}N_2 \frac{m_{s,{\rm (eff)}}^2}{3H^2}(N_2)\right] ,
\ee
where we have assumed that $\chi$ has effectively vanished by the time it becomes light, that we denote $N_\ell$. This allows one to compute the second moment of $\chi$
\begin{align}
\left\langle \chi^2 \right\rangle\left(N\right) & = 
\int_{N_\ell}^N \mathrm{d} N_1 \frac{H^2(N_1)}{4\pi^2} \exp\left[-\frac{2}{3}\int_{N_1}^{N}\mathrm{d}N_2 \frac{m_{s,{\rm (eff)}}^2}{H^2}(N_2)\right]\\
  & =  \int_{N_\ell}^N \mathrm{d} N_1 \frac{H^2(N_1)}{4\pi^2} \exp\left\lbrace -\frac{2}{3}\int_{N_1}^{N}\mathrm{d}N_2 \left[ \frac{m^2}{H^2\left(N_2\right)}-4\left(\frac{M_{{}_\mathrm{Pl}}}{M}\right)^2 \epsilon_1\left(N_2\right)\right]\right\rbrace
 , \label{chi2}
\end{align}
where, in the second line, Eq.~(\ref{eq:ms2_2}) has been employed. For any given model, the integral \eqref{chi2} can be evaluated numerically. However, one can obtain a useful analytical insight by using the slow-roll approximation, leading to the following expression for $\langle\chi^2_\mathrm{c}\rangle\equiv  \langle\chi^2\rangle(N_\mathrm{c})$ (see Appendix for details):
\be
\left\langle \chi^2_\mathrm{c} \right\rangle \simeq  \left( \frac{H_\mathrm{c}}{2\pi} \right)^2 \left\lbrace \frac{1}{2}\sqrt{3 \pi \DeltaN}\mathrm{erf}\!\left[\sqrt{\DeltaN/3} \right]-3 \epsilon_{1,\mathrm{c}} \DeltaN \left[ e^{-\DeltaN/3}-1\right] \right\rbrace ,
\label{eq:ktd1}
\ee
where
\be
\DeltaN \equiv N_\mathrm{c}-N_\ell  \simeq  \left(\frac{H_\mathrm{c}}{m}\right)^2 \frac{1}{\epsilon_{2,\mathrm{c}}-2 \epsilon_{1,\mathrm{c}}}\,
\label{deltaN-light}
\ee
is the number of $e$-folds elapsed in the light but stabilised phase. Expanding Eq.~(\ref{eq:ktd1}), we obtain the two limiting behaviours
\be
\left\langle \chi^2_\mathrm{c} \right\rangle \simeq \left\{ 
\begin{array}{ll} 
\left( \dfrac{H_\mathrm{c}}{2\pi} \right)^2 \DeltaN \quad \textrm{for}\ \DeltaN \ll 1\,, \\
\left( \dfrac{H_\mathrm{c}}{2\pi} \right)^2 \left( \frac{1}{2}\sqrt{3 \pi \DeltaN} +3 \epsilon_{1,\mathrm{c}} \DeltaN  \right) \quad \textrm{for}\ \DeltaN \gg 1\,.
 \end{array} \right.
 \label{eq:ktd3}
\ee
The first result holds when the duration of the phase during which $\chi$ is light but stabilised is short 
(in terms of the number of e-folds it lasts for), and its agreement with the expected free diffusion limit (see \textit{e.g.}~\cite{Hardwick:2017fjo}) provides a good consistency check. In that case, the typical field displacement is suppressed compared to $H_{\mathrm{c}}$. The second result corresponds to when the duration of the light stable phase is long. It is always suppressed compared to the free diffusion limit, which can be seen as the consequence of the small (but positive) mass of the field reducing its growth. In that case, the typical field displacement is nonetheless much larger than $H_{\mathrm{c}}$. 
{It should be noted that in deriving \Eq{eq:ktd3}, we neglected the time dependence of $\chi$, \ie the possibility that bending becomes important early in the dynamics of geometrical destabilisation. 
We shall comment later on the validity of this assumption.}
\\

\begin{figure}
\begin{center}
\includegraphics*[width=6.7cm]{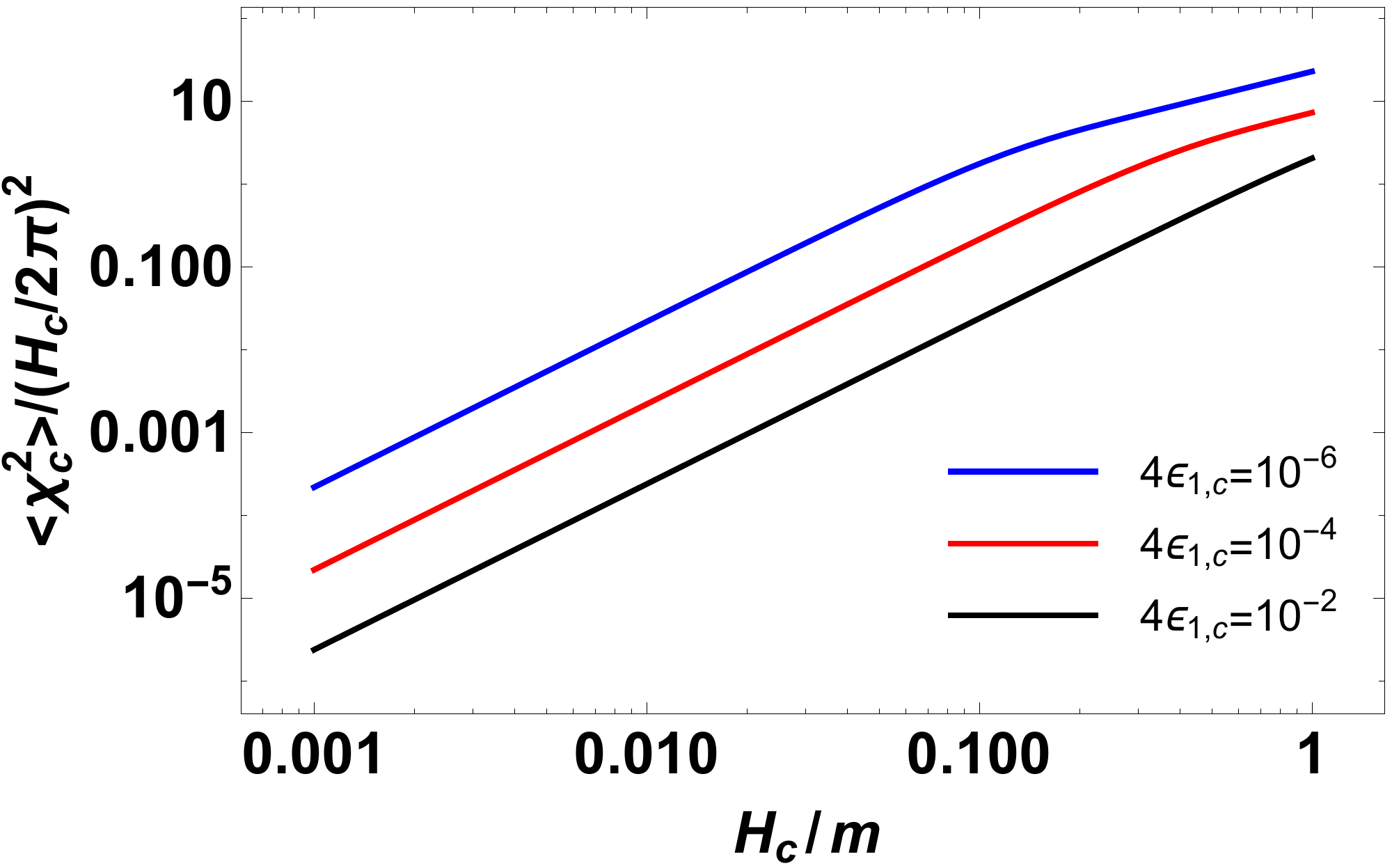}
\hspace{1cm}
\includegraphics*[width=6.7cm]{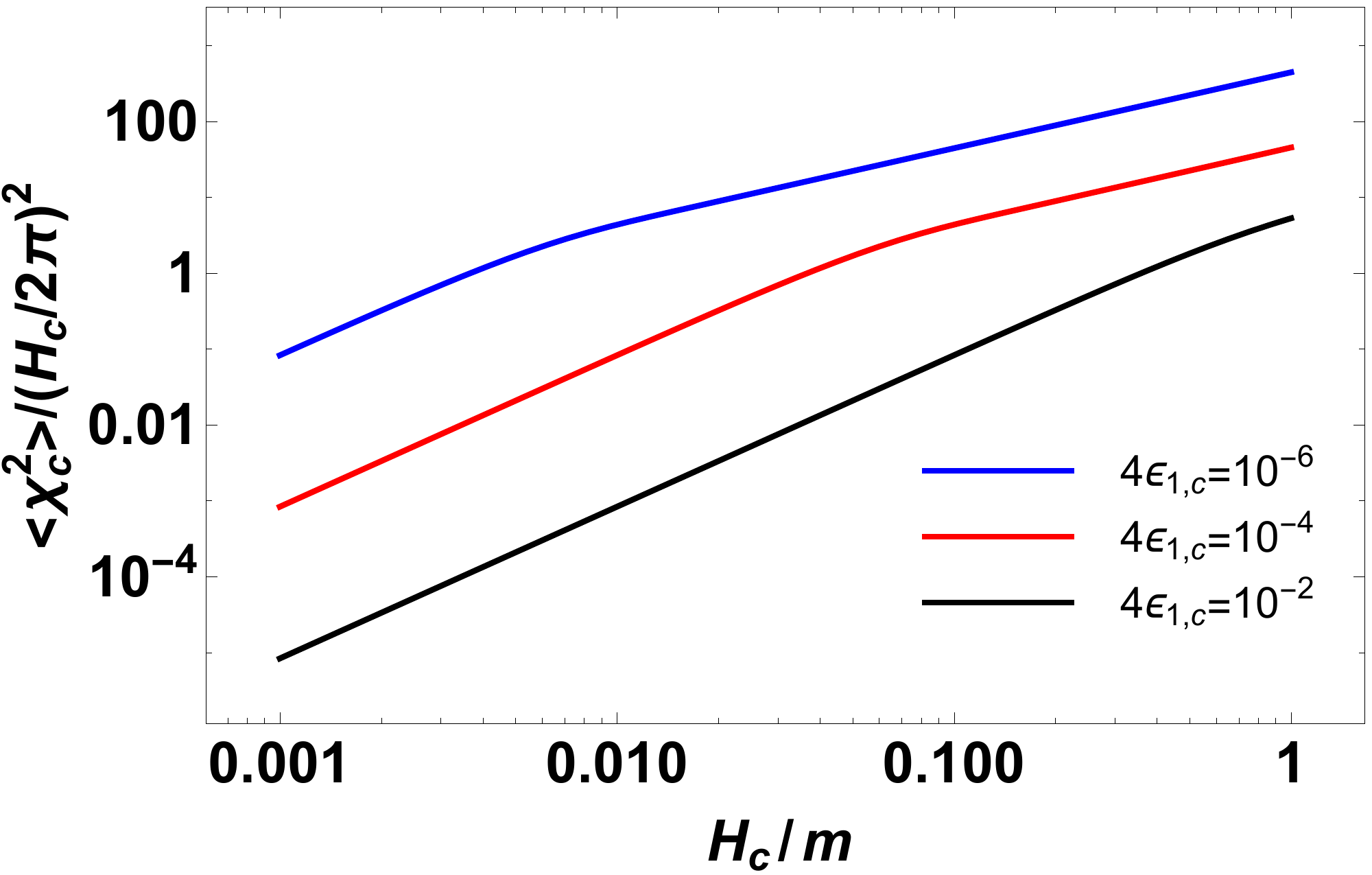}
\caption{Spectator variance $\langle\chi_\mathrm{c}^2\rangle$ at the beginning of geometrical stabilisation, computed with the slow-roll estimate \eqref{eq:ktd1}, as a function of the ratio between the Hubble parameter and the spectator mass, and for different values of $\epsilon_{1, {\rm c}}$. Left:  Starobinsky potential~(\ref{eq:Starobinsky}). Right: monomial potential $V(\phi) \propto \phi$. 
\label{fig:initial-displacement}}
\end{center}
\end{figure}

Specifying a model can provide a relation between $\epsilon_{1,\mathrm{c}}$ and $\epsilon_{2,\mathrm{c}}$,
which in turn allows one to express $\langle \chi^2_\mathrm{c} \rangle$ in terms of the parameters describing the geometrical destabilisation. We use this in \Fig{fig:initial-displacement} to display the spectator variance $\langle\chi_\mathrm{c}^2\rangle$ at the beginning of geometrical stabilisation for various combinations of parameters, for the Starobinsky potential~(\ref{eq:Starobinsky})  (left), and for the monomial potential $V(\phi) \propto \phi$ (right), which correspond respectively to $\epsilon_{2,\mathrm{c}} = \frac{4}{\sqrt{3}} \sqrt{\epsilon_{1,\mathrm{c}}}$ and $\epsilon_{2,\mathrm{c}} =4\epsilon_{1,\mathrm{c}}$ (see appendix \ref{sec:app1}). In each case, the two limiting behaviours \eqref{eq:ktd3} are clearly visible, as well as the dependence on parameters: the smaller the $\epsilon_{1,\mathrm{c}}$ parameter (and hence $\epsilon_{2,\mathrm{c}}$ in these models), the larger the number of $e$-folds spent in the light stabilised phase, and hence the larger the variance; and the larger the hierarchy $m/H_\mathrm{c}$, the shorter the duration of the stabilised phase, and the smaller the typical field displacement.

The validity of the slow-roll estimate \eqref{eq:ktd1} is discussed in detail in appendix \ref{sec:app1}. It becomes inaccurate only if $\epsilon_{2,\mathrm{c}} \simeq 2\epsilon_{1,\mathrm{c}}$, and more precisely if $\epsilon_{2,\mathrm{c}}/(2\epsilon_{1,\mathrm{c}})-1 ={\cal O}(H_\mathrm{c}/m)^2$, which corresponds to a very limited region of models and parameter space. In the appendix, we consider the case of monomial potentials $V(\phi) \propto \phi^p$, which are amenable to exact computations, and show that the slow-roll estimate \eqref{eq:ktd1} indeed becomes inaccurate only for $p$ very close to $2$, corresponding to the limiting case $\epsilon_{2}=2\epsilon_{1}$.

\subsection{Kinematical backreaction in geometrical destabilisation}
\label{sec:kin}

destabilisation of the inflationary trajectory in the $\chi$ direction can significantly affect the motion of the inflaton field $\phi$.
Indeed, writing down the equation of motion (\ref{eq:scfi_eom}) in the context of the action (\ref{minimal_two}), we obtain
\be
\label{eq:phi_eom}
\ddot{\phi} +3H\dot{\phi}+2b'\dot{\phi}\dot{\chi}+e^{-2b}V_\phi = 0\, .
\ee
Even if the field-space motion predominantly takes place in the $\phi$ direction, such that we can neglect the third term in Eq.~(\ref{eq:phi_eom}), the
fact that $\chi$ increases during geometrical destabilisation (due to the amplification of long-wavelength $\chi$-modes) can effectively reduce the slope of the potential through the inverse
of the field-space metric in the fourth term in (\ref{eq:phi_eom}). This can slow down the field $\phi$ so that the slow-roll parameter $\epsilon_1$ is reduced
and the instability condition is no longer satisfied. 

With $\R=-2(b''+{b'}^2)$ for the action (\ref{minimal_two}), the geometrical contribution to $m^{2}_{s {\rm (eff)}}$ given by the third term in \Eq{ms2} can be expressed as:
\be
\label{eq:ms2_geo}
    \epsilon_1  \R \M^2 \simeq - \left(\frac{\dot{\phi}}{H}\right)^2 e^{2b} \left(b''+{b'}^2\right)  ,
\ee
where we dropped the contribution to $\epsilon_1$ proportional to $\dot{\chi}^2$, and the slow-roll equation for the inflaton reads
\be
\label{eq:phisr}
\dot\phi \simeq - \frac{e^{-2b}V_\phi}{3H} \, . 
\ee
Substituting \Eq{eq:phisr} into \Eq{eq:ms2_geo}, we obtain
\be
\label{eq:e24}
\epsilon_1  \R \M^2 = -\left( \frac{V_\phi}{3H^2} \right)^2 e^{-2b}\left(b''+{b'}^2\right)  = - \left( \frac{V_\phi}{3H^2} \right)^2e^{-3b} \frac{\mathrm{d}^2e^b}{\mathrm{d}\chi^2} \, .
\ee
We find this form of the geometrical contribution to the effective mass $m_{s\mathrm{(eff)}}^2$ particularly useful, because neither $V_\phi$ nor $H$ changes significantly during
geometrical destabilisation and practically the entire $\chi$-dependence is encoded into the function $e^{-3b} \frac{\mathrm{d}^2e^b}{\mathrm{d}\chi^2} $.
Within the minimal model (\ref{minimal}), we have $e^{2b}=1+2\chi^2/M^2$, which leads to
\be
\label{eq:Rfs:minimal:model}
e^{-2b}\left(b''+{b'}^2\right) = \frac{2}{M^2}\left( 1+\frac{2\chi^2}{M^2}\right)^{-3}  .
\ee
As $\chi$ grows, the negative contribution to $m^{2}_{s {\rm (eff)}}$ becomes suppressed by a factor $\sim30$ as $\chi$ increases from zero to $M$. 
This suppression is a joint effect of the slowing down of $\phi$ due to the non-canonical kinetic term and a suppression of the field-space curvature
away from $\chi=0$. 
We therefore conclude that in the minimal
realisation of geometrical destabilisation, this backreaction effectively shuts off the instability. We refer to this backreaction as ``kinematical'', because it does not
rely on any particular detail of the inflationary dynamics and originates purely from the non-canonical form of the kinetic term for the inflaton.
This conclusion remains valid also for the generalised model~(\ref{minimal_two}), as long as the curvature of the field space at $\chi=0$ is
$R_\mathrm{fs}=-4/M^2$
and
the function $e^{2b(\chi)}$ that enters the non-canonical kinetic term for the inflaton
can be reliably expanded as:
\be
e^{2b(\chi)} = 1+\frac{2\chi^2}{M^2} + \frac{c_3}{3!}\left(\frac{\chi}{M} \right)^3 +  \frac{c_4}{4!}\left(\frac{\chi}{M} \right)^4 +\cdots .
\ee
As the inflaton $\phi$ rolls, the slope of the potential, $V_\phi$, becomes steeper, so $m^{2}_{s {\rm (eff)}}$ becomes negative and large, until kinematical backreaction becomes effective. 
At that time, the inflationary trajectory may start turning, so we also include the bending term that we estimate as:
\be
\label{eq:etaperp1}
3\eta_\perp^2H^2 = 3 \left( \frac{m^2\chi}{e^b\dot{\phi}H}\right)^2 H^2 \simeq \frac{6m^2\chi^2}{M^2} \, ,
\ee
where in the first step we used Eq.~(\ref{etaperpdefinition}) and identified the $\chi$ direction as perpendicular to the inflationary trajectory, and in the second step we 
expanded the result around the critical time to the lowest order and used Eq.~(\ref{eq:epsilonc}).
Expanding Eq.~(\ref{eq:e24}) around the critical time and combining all contributions to the effective mass,
we can finally express $m^{2}_{s {\rm (eff)}}$ as
\be
\label{eq:mmax1}
m^{2}_{s {\rm (eff)}} = m^2\left[-\left(\epsilon_{2,\mathrm{c}}-2\epsilon_{1,\mathrm{c}}\right)\left(N-N_{\mathrm{c}}\right)-\frac{c_3}{4}\frac{\chi}{M}+ \left(12-\frac{c_4}{8}\right)\frac{\chi^2}{M^2}\right] .
\ee
 It should be stressed that Eq.~(\ref{eq:mmax1}) results from two expansions: the first term is obtained by expanding $V_\phi$ and $H$ in Eq.~(\ref{eq:e24}) in time and writing the result in terms of the slow-roll parameters $\epsilon_{1,\mathrm{c}}$
 and $\epsilon_{2,\mathrm{c}}$ calculated at the critical time, while the remaining terms are obtained by expanding $e^{-3b} \frac{\mathrm{d}^2e^b}{\mathrm{d}\chi^2}$ in Eq.~(\ref{eq:e24}) in $\chi$ and using Eq.~(\ref{eq:etaperp1}); the minimal model corresponds to $c_3=c_4=0$.
 We can see that there is a positive contribution to $m^{2}_{s {\rm (eff)}}$ that tends to shut off the instability as $\chi$ grows. The precise value of $\chi$ at which
 this happens is model dependent, but Eq.~(\ref{eq:mmax1}) strongly suggests that geometrical destabilisation is over no later than when $\chi$ reaches $M$. We can define the maximal scale
 describing the instability as $m_\mathrm{max}^2\equiv \mathrm{max}\left[ -m^{2}_{s {\rm (eff)}}\right]$. As we shall see shortly, the precise value of $m_\mathrm{max}^2$ can be quite sensitive to the parameters of the model.

We illustrate the discussion presented here by solving numerically the equations of motion for the homogeneous fields in the minimal realisation of geometrical destabilisation using the Starobinsky potential  (\ref{eq:Starobinsky}) with $\Lambda=2\times10^{-6}\M$ for the inflaton. We consider two sets of parameters,
\be
\left( \frac{m}{H_\mathrm{c}},\frac{\M}{M} \right) = (10,10^3)\,\,\textrm{and}\,\,(10^2,10^4)\,,
\ee
corresponding to the same value $\epsilon_{1,\mathrm{c}}=2.5\times10^{-5}$. 
The initial value of the spectator field is set according to our discussion in \Sec{sec:ini}, {\it{i.e.}} taking $\chi=\sqrt{\langle\chi_\mathrm{c}^2\rangle}$ 
calculated from Eq.~(\ref{eq:ktd3}).

\begin{figure}
\begin{center}
\includegraphics*[width=7.4cm]{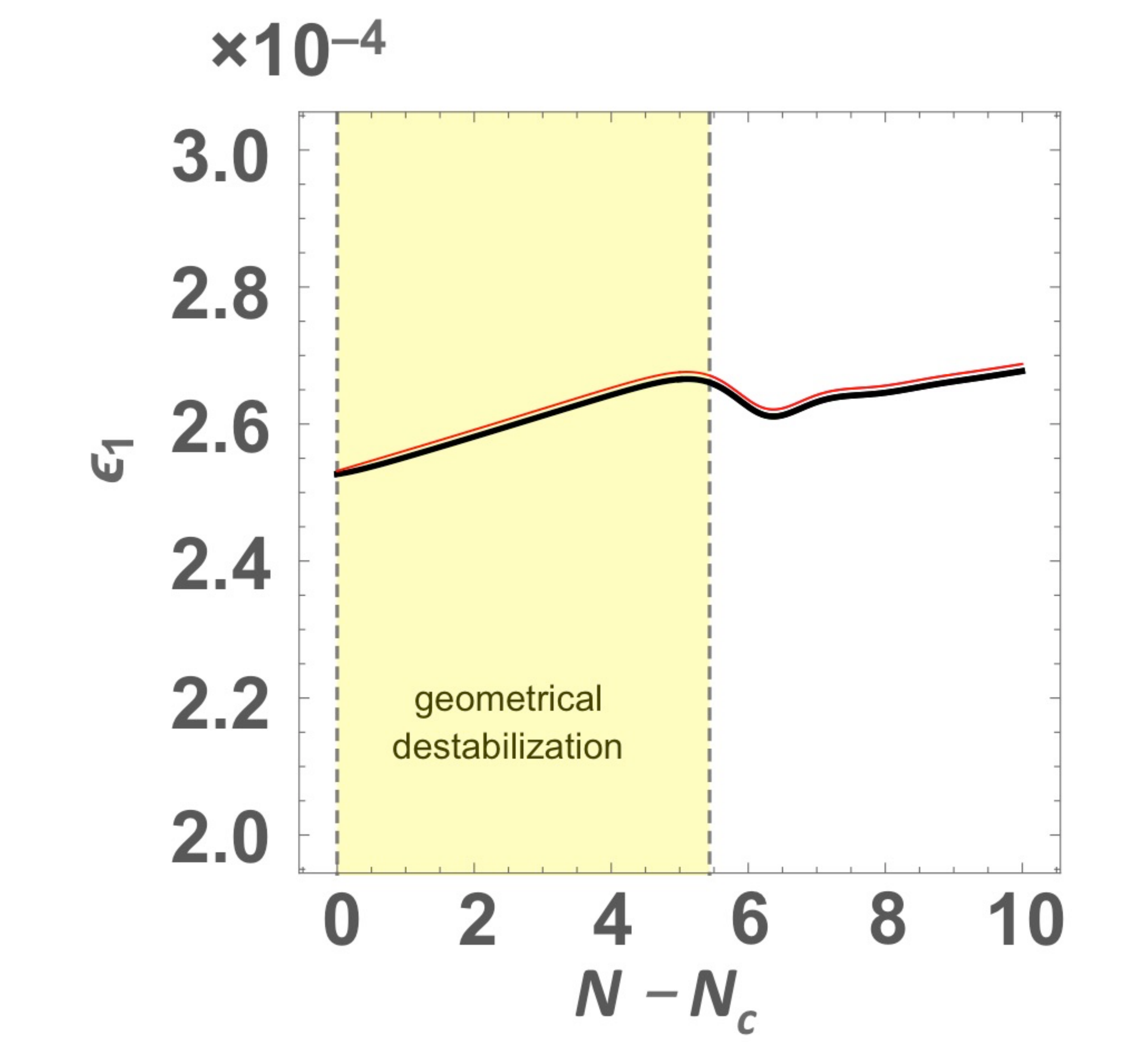}
\hspace{1cm}
\includegraphics*[width=7.4cm]{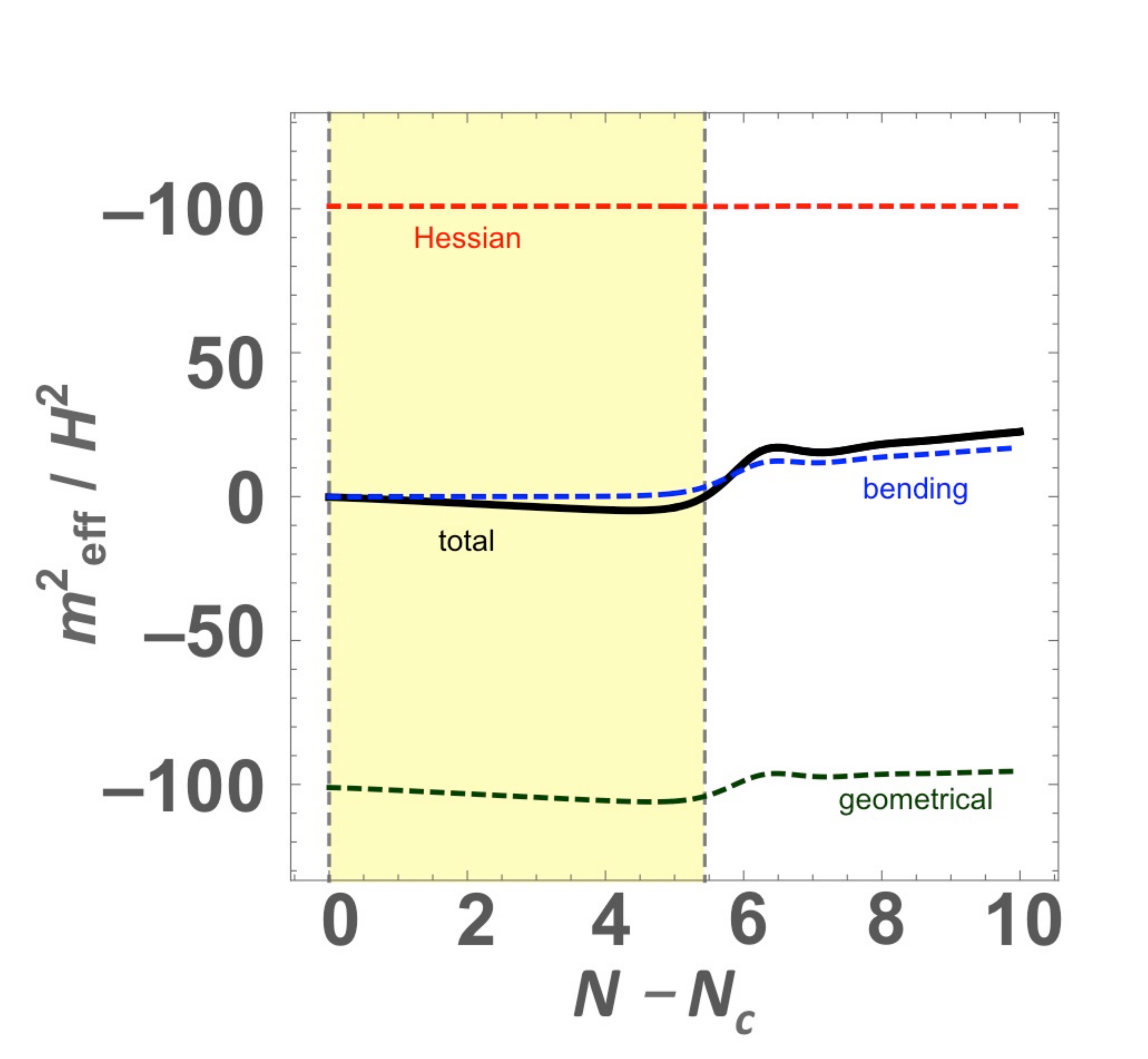}
\caption{{Background evolution for the minimal model \eqref{minimal}, with the Starobinsky potential \eqref{eq:Starobinsky} for the inflaton, \mbox{$\Lambda=2\times10^{-6}\M$}, and the parameters $\left( \frac{m}{H_\mathrm{c}},\frac{\M}{M} \right) = (10,10^3)$ for the sector of geometrical destabilisation. The initial value of the field $\chi$ is $\sqrt{\langle\chi_\mathrm{c}^2\rangle}$, computed with \Eq{eq:ktd1}, and its initial velocity is zero.} Left panel: Evolution of the slow-roll parameter $\epsilon_1$ during and after geometrical destabilisation. $N_{\mathrm{c}}$ is the reference number of $e$-folds at which geometrical destabilisation begins. The thick black line corresponds to the exact numerical calculation, {while the thin red line uses the slow-roll result following from Eq.~(\ref{eq:phisr}), computed on the exact solution}. The shaded area marks the period of geometrical destabilisation, {defined as $m^{2}_{s {\rm (eff)}}<0$}; 
when it ends, we have $\chi_\mathrm{end}/M\approx0.077$, which corresponds to $e^{2b}\approx1.012$.
Right panel: Evolution of $m^{2}_{s {\rm (eff)}}/H^2$ (black line) and its three contributions: Hessian, bending and geometrical terms.} 
\label{f:kb}
\end{center}
\end{figure}

\begin{figure}
\begin{center}
\includegraphics*[width=7.4cm]{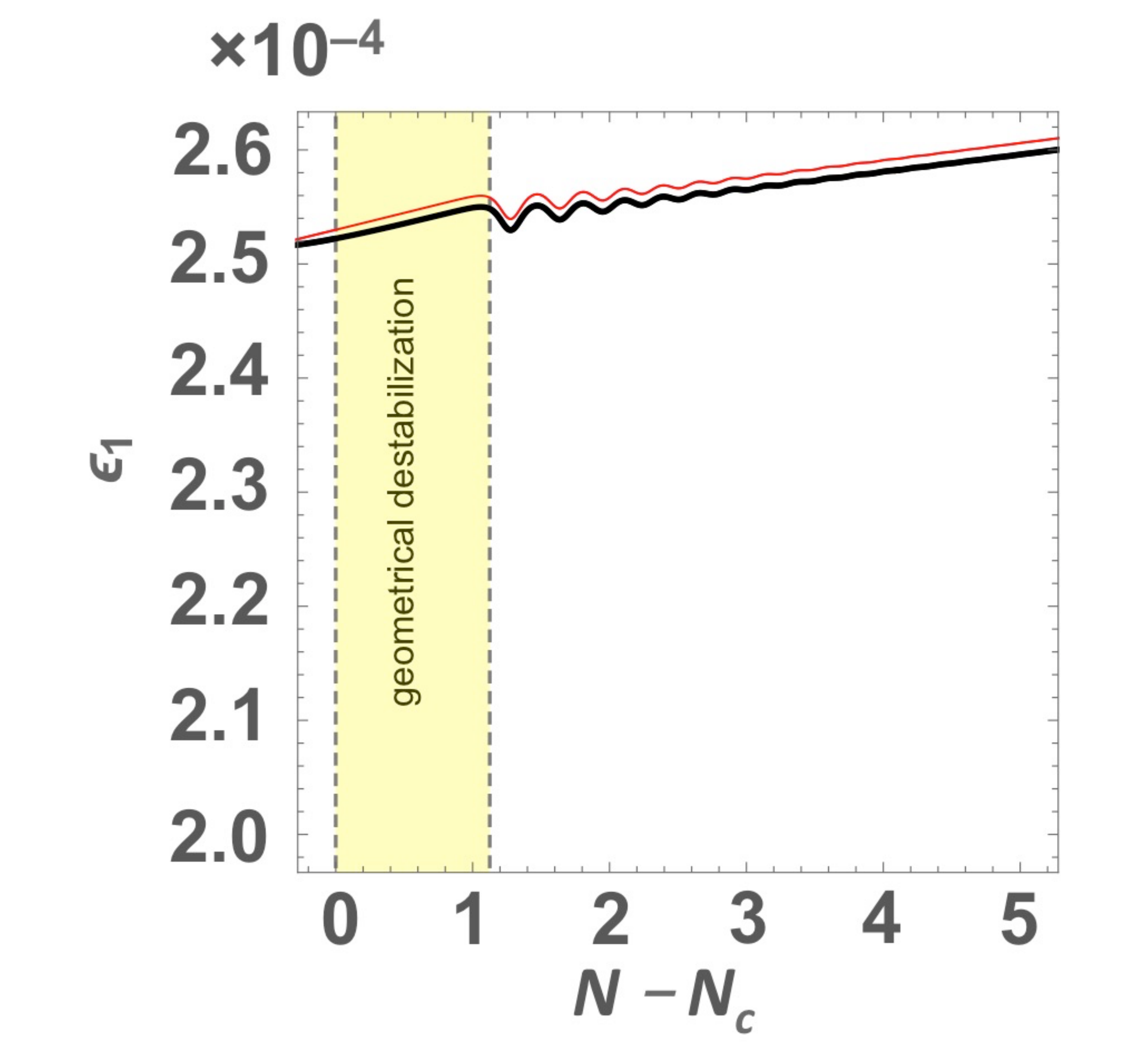}
\hspace{1cm}
\includegraphics*[width=7.4cm]{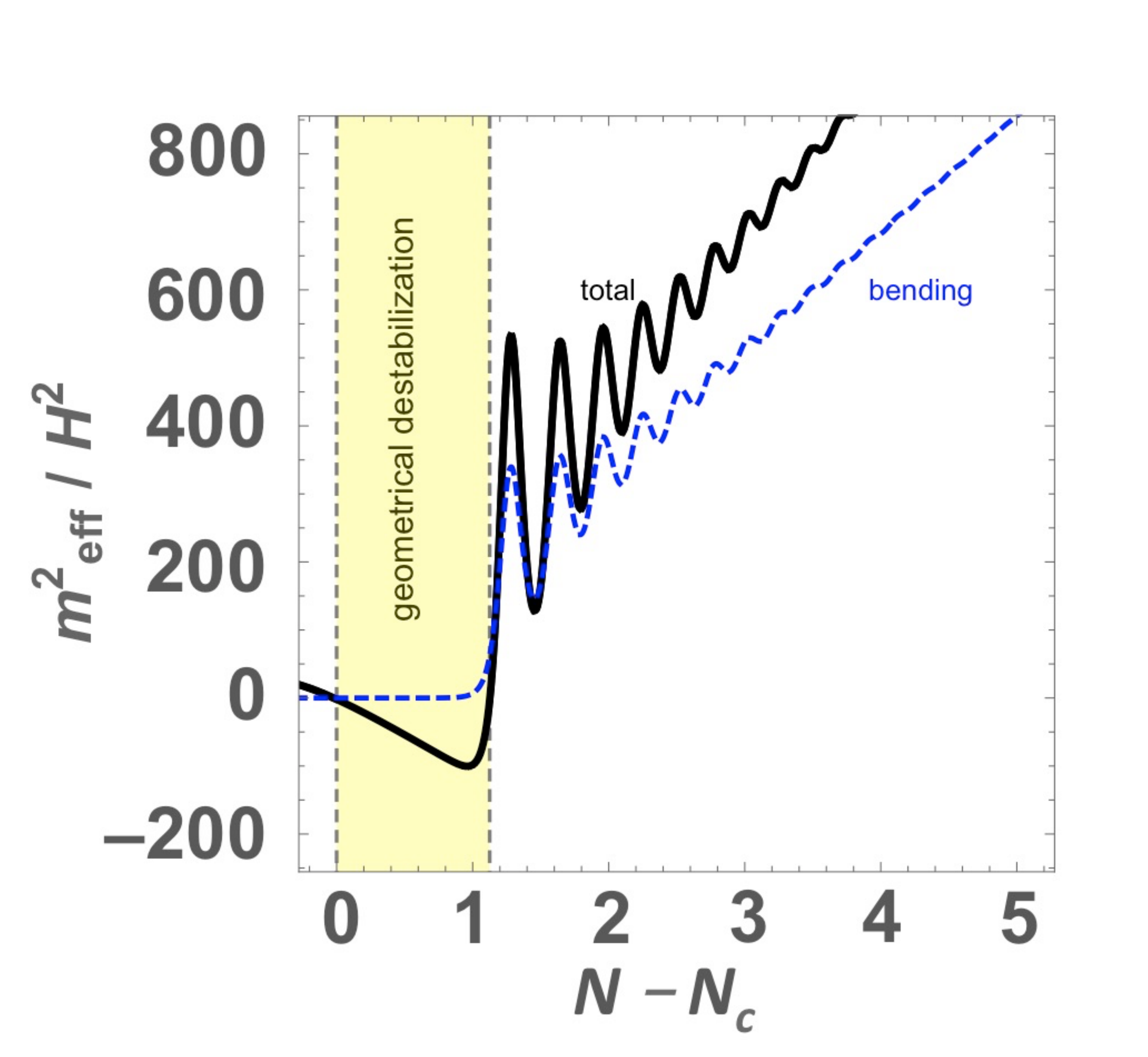}
\caption{{Same as in \Fig{f:kb}, for the parameters $\left( \frac{m}{H_\mathrm{c}},\frac{\M}{M} \right) = (10^2,10^4)$. Geometrical destabilisation ends when $\chi_\mathrm{end}/M\approx0.032$, which corresponds to $e^{2b}\approx1.0021$.}
\label{f:kb2}}
\end{center}
\end{figure}

The results for the two sets of parameters are shown in \Figs{f:kb} and \ref{f:kb2}, respectively.
We display the time evolution
of the slow-roll parameter $\epsilon_1$, computed either exactly or with the slow-roll approximation (\ref{eq:phisr}),
as well as the evolution of the various contributions to the effective mass of the spectator field. 
At the beginning, the first term in \Eq{eq:mmax1} dominates and $m_{s\mathrm{(eff)}}^2$ decreases linearly below zero. Towards the end of geometrical destabilisation, the exponential growth of $\chi$
 leads to the diminishing of $\epsilon_1$, increases $|\eta_\perp|$ and the positive terms in \Eq{eq:mmax1}, and, thereby, shuts off the instability.
Note that 
the cancellation between the Hessian and the geometrical contribution to $m_{s,{\rm (eff)}}^2$ is quite delicate and a decrease in $\epsilon_1$ by a few per cent is sufficient to terminate
geometrical destabilisation. The value of the field $\chi_\mathrm{end}$ corresponding to the end of geometrical destabilisation can be inferred from (\ref{eq:mmax1}):
for the minimal realisation with $c_3=c_4=0$, we find
\be
\label{eq:chiend}
\frac{\chi_\mathrm{end}}{M} = \sqrt{\frac{(2\epsilon_{1,\mathrm{c}}-\epsilon_{2,\mathrm{c}})(N_\mathrm{end}-N_\mathrm{c})}{12}} \, ,
\ee
where we  
denoted by $N_\mathrm{end}$ the number of e-folds corresponding to the end of geometrical destabilisation.
\Eq{eq:chiend} is in a very good agreement with our numerical results, to the per cent level. We also find that $m_\mathrm{max}^2$ is roughly equal to $5H_\mathrm{c}^2$ and $100H_\mathrm{c}^2$,
respectively for the first and second parameter choice.
When geometrical destabilisation is over, the bending term becomes sizeable and inflation continues along a new, stable, sidetracked trajectory \cite{GD2}.

Let us note that our examples are consistent with \Ref{Cicoli:2018ccr}, in which inflationary trajectories with different initial displacements from $\chi=0$
were analysed in the context of the minimal realisation of geometrical destabilisation; it was found that they generically deviate from $\chi=0$ and eventually approach a stable attractor solution.

\subsection{Universality of kinematical backreaction}
\label{sec:uni}

We shall now argue that the kinematical backreaction effects found in our numerical examples are, in fact, unavoidable. It follows from the discussion presented in \Sec{sec:kin} that the instability could be present for a long time, if the geometrical contribution (\ref{eq:e24}) to the mass term (\ref{ms2}) was negative
and sufficiently large. 
We can rewrite the latter postulate in terms of a constrain on the form of the non-canonical kinetic term for the inflaton in \Eq{minimal_two}.
Adopting the notation $f(\chi)=e^{b(\chi)}$, we can ask if there exists a nonsingular function $f$ satisfying:
\be
\label{eq:inif}
f(0)=1\,,\qquad f'(0)=0\,,\qquad f''(0) = \frac{4}{M^2}
\ee
together with
\be
\label{eq:dif_f1}
f^{-3}f'' = g\,,
\ee
where $g$ is a positive function such that $\mathrm{inf}\,g=c>0$ for some constant $c$.
In this way, the geometrical contribution to the effective mass $m_{s {\rm (eff)}}^2$ given in \Eq{eq:e24} is always smaller than $-c(V_\phi/3H)^2$; with a sufficiently large $c$, we would like to
prevent kinematical backreaction from ending geometrical destabilisation.
Here we assume that the mass of the spectator is constant and that the slope of the inflationary potential $|V_\phi|$ does not decrease during geometrical destabilisation, 
so a single value of $c$ ensures that the destabilisation condition is satisfied at all times. We also assume that the non-negative bending contribution to $m_{s,{\rm (eff)}}^2$
is negligible; it is a conservative assumption, because a sizeable bending contribution tends to end geometrical destabilisation before the bounds that we estimate here are saturated.

Let the function $\tilde{f}$ be a solution to the following equation:
\be
\label{eq:dif_f2}
\tilde{f}^{-3}\tilde{f}'' = c\, ,
\ee
with initial conditions $\tilde{f}(0)=1$ and $\tilde{f}'(0)=0$. Both $f$ and $\tilde{f}$ are increasing positive functions. Their initial conditions and the inequality $g(\chi)\geq c>0$
imply that $f(\chi)\geq \tilde{f}(\chi)$ in the common domain of the solutions of \Eqs{eq:dif_f1} and~(\ref{eq:dif_f2}). For the latter equation, there exists a first
integral
\be
\label{eq:fenergy}
E = \frac{1}{2}\tilde{f}'^2 - \frac{c}{4}\tilde{f}^4 \, .
\ee
Applying initial conditions, we find $E=-\frac{c}{4}$. Separating variables in (\ref{eq:fenergy}), we obtain
\be
\mathrm{d}\chi = \sqrt{\frac{2}{c}} \frac{\mathrm{d}\tilde{f}}{\sqrt{\tilde{f}^4-1}} \, .
\ee
Integrating within appropriate limits, we find:
\be
\label{eq:gkk1}
\chi_\mathrm{max} = \sqrt{\frac{2}{c}} \int_1^\infty \frac{\mathrm{d}\tilde{f}}{\sqrt{\tilde{f}^4-1}} = \sqrt{\frac{2\pi}{c}} \frac{\Gamma\left(\frac{5}{4}\right)}{\Gamma\left(\frac{3}{4}\right)} \approx \frac{1.854}{\sqrt{c}} \, .
\ee
We conclude that $\tilde{f}$ becomes singular for a finite value of $\chi_\mathrm{max}$, hence $f$ also becomes singular at some point between zero and $\chi_\mathrm{max}$. 
For $|\chi|\leq \chi_\mathrm{max}$, the function $\tilde{f}$ is completely specified by \Eq{eq:dif_f2} and the initial conditions $\tilde{f}(0)=f(0)=1$ and $\tilde{f}(0)=f(0)=0$.
If we additionally demand that $\tilde{f}''(0)=f''(0)=4/M^2$, this determines the value of $c$ in \Eq{eq:dif_f2} as $c=4/M^2$, which implies 
$\chi_\mathrm{max}\approx0.927M$. 

With \Eq{eq:dif_f1}, one finds that $\R=-2 f^{''}/f$ is such that $|R_\mathrm{fs}| \geq 2 c f^2$ and hence the singularity described above is a genuine singularity of the field-space curvature at $\chi\sim M$.
This introduces
a strict upper bound on the value of the field $\chi$, meaning that its amplitude cannot grow arbitrarily large, which is at odds with our starting assumption.
This is because divergent $b(\chi)$ translates 
through \Eq{eq:phisr} to $\epsilon_1\to0$, as both $V_\phi$ and $H$ remain approximately constant. Consequently, the vanilla picture of geometrical destabilisation painted in \Sec{sec:kin} must break down before $\chi$ reaches $M$ in the case of a field space with curvature singularity. Discarding this situation, whose study lies outside the scope of this work, we conclude that the kinematical backreaction is unavoidable in regular field spaces.

One can further ask whether, during the time in which geometrical destabilisation is active, the exponential growth of the perturbations of $\chi$
is capable of terminating inflation. We shall address this question in the following section.

\section{Linear analysis of perturbations during geometrical destabilisation}
 \label{sec:linana}
 
Geometrical destabilisation is certainly not the first example of an instability of the inflationary dynamics that may strongly affect the predictions of inflation. Hybrid inflation~\cite{hybrid0,hybrid00,hybrid} is a very well-known and thoroughly studied class of models with an instability that ultimately terminates inflation. 
Many papers, including \Refs{Lyth:2010zq,Lyth:2012yp}, dealt with the problem
 of the evolution of perturbations during the instability. The setup of geometrical destabilisation differs from hybrid 
 inflation by the fact that the instability is not encoded in the potential, but it is a purely dynamical, velocity-dependent
 effect that stems from the non-zero curvature of the field space. However, in both cases there is a scalar field, whose
 fluctuations acquire a negative mass squared, so the methods and techniques that have been successfully applied 
 to hybrid inflation can be also used for geometrical destabilisation.

 \subsection{Evolution of perturbations}
 \label{end?}
 
In this section, we closely follow \Refs{Lyth:2010zq,Lyth:2012yp} to determine if geometrical destabilisation can affect the inflationary dynamics to the extent that the perturbative treatment breaks down and inflation can be terminated.Therefore, we make the working assumption that this is indeed the case and study the consistency of this ansatz with inflationary predictions. 
 
In \Sec{sec:kin}, it was shown that geometrical destabilisation can proceed along two regimes: either it is long and mild, \ie it lasts for a few $e$-folds before kinematical backreaction kicks in, and $m_{\mathrm{max}}$ is not much larger than $H_{\mathrm{c}}$ (this is the case displayed in \Fig{f:kb}), either it is short and violent, \ie it lasts for $\lesssim {\cal O}(1)$ $e$-folds, and $m_{\mathrm{max}}$ grows much larger than $H_{\mathrm{c}}$ (see \Fig{f:kb2}). In this section, we consider the second regime of parameter space, since in the first regime, a mild and prolonged amplification of perturbations is more likely to induce large renormalisations of the super-Hubble effective ``background'' than to terminate inflation. We therefore neglect the expansion of the universe and assume $m_{\mathrm{max}}\gg H_{\mathrm{c}}$ in what follows. 
 
With this assumption, we can initially disregard the kinematical backreaction, and perform the perturbative expansion about $\chi=0$ as in subsection \ref{sec:kin}. This is because, as will be shown below, a short and violent geometrical destabilisation is such that $\DeltaN \ll 1$, so $\langle \chi_{\mathrm{c}}^2 \rangle \simeq [H_{\mathrm{c}}/(2\pi)]^2 \DeltaN$ according to \Eq{eq:ktd3}, and
 \be
 \left\vert \eta_{\perp\mathrm{c}} \right\vert = \sqrt{\frac{\DeltaN}{2\pi^2}} \frac{m}{M} 
 \ee
 according to \Eq{eq:etaperp1}. By considering realistic values of $m$ and $M$ such that $M \gtrsim m$, we deduce that $\vert \eta_{\perp\mathrm{c}} \vert \ll 1$, which means that the bulk of geometrical destabilisation happens before the inflationary trajectory starts turning\footnote{{On the other hand, with a long destabilisation phase, $\DeltaN \gg 1$, it follows from \Eqs{eq:ktd3} and~\eqref{eq:etaperp1} that $\eta_\perp$ may not be negligible and the turning contribution  may be important in \Eq{ms2}. In that case, the reasoning that led to \Eq{eq:ktd3} is not self-consistent; however, this corresponds to a slow and mild destabilisation.}}.
  
Making use of the expansion~\eqref{eq:mmax1} of the effective mass around the critical time, denoted as $t_{{\rm c}}=0$, and with $\chi=0$, we thus write
 \be
 \label{eq:ilawa2}
m^{2}_{s {\rm (eff)}} = -\mu^3t \, 
\ee
with the mass scale 
\be
\mu^3 = m^2 (\epsilon_{2,\mathrm{c}}-2\epsilon_{1,\mathrm{c}})H_{\rm c}\,.
\label{mu}
\ee
Of course, \Eq{eq:ilawa2} is valid only in a certain time interval around the beginning of geometrical destabilisation and it should not be applied at too early times. In particular, when the scales observable in the Cosmic Microwave Background (CMB) leave the Hubble radius, we expect the geometrical contribution to $m^{2}_{s {\rm (eff)}}$ to be small and $m^{2}_{s {\rm (eff)}}\sim m^2$. 

We note that the mass scale $\mu$ is related to the duration~\refeq{deltaN-light} of the light stabilised phase  preceding the geometrical destabilisation as $\mu^3/H_{{\rm c}}^3=1/\DeltaN$, and that the linear expansion \eqref{eq:ilawa2} can also be rewritten as 
\be
\frac{m^{2}_{s {\rm (eff)}}}{H_{{\rm c}}^2}=-\frac{N-N_{{\rm c}}}{\DeltaN}\,.
\ee
Following the discussion above, we assume that the timescale of the instability is short compared to the timescale of expansion of the universe, \textit{i.e.} $\mu \gg H_{\mathrm{c}}$, or equivalently that it takes a small fraction of $e$-folds $\DeltaN$ to reach $m^{2}_{s {\rm (eff)}}=-H_{{\rm c}}^2$ from the critical time onwards (this confirms the validity of the assumption $\DeltaN\ll 1$ made above). We note that the two examples presented in \Sec{sec:kin} correspond to $\mu^3/H_{{\rm c}}^3=1.15$ and $115$, respectively, so the second case indeed lies within the realm of approximations considered in this section.

Neglecting the expansion of the Universe, the equation of motion for the Fourier component of the unstable
field $\chi$ reads:
\be
\label{e22}
\ddot{\chi}_{\bm{k}} + \left[k^2+m^{2}_{s {\rm (eff)}}\right] \chi_{\bm{k}}  = 0 \, .
\ee
Introducing $\tau=\mu t$ and $x=\tau-\frac{k^2}{\mu^2}$, we can rewrite \Eq{e22} as
\be
\label{e23}
\frac{\mathrm{d}^2\chi_{\bm{k}}}{\mathrm{d}x^2} - x\chi_{\bm{k}} = 0 \, .
\ee
The solution of Eq.~(\ref{e23}) can be written as:
\be
\label{eq:airy}
\chi_{\bm{k}} = C_1\,\mathrm{Ai}(x)+C_2\,\mathrm{Bi}(x)\,,
\ee
where Ai and Bi are
the Airy functions.
The coefficients $C_1$ and $C_2$ of this linear combination can be determined form the requirement that for early
times, $x\ll-1$, when the mode evolution is adiabatic, one has a positive-frequency
and properly normalised solution:
\be
\chi_{\bm{k}} = \frac{1}{\sqrt{2\omega_{\bm{k}}}}e^{-\imath \int^t \omega_{\bm{k}}\mathrm{d}t} \sim \frac{1}{\sqrt{2\mu}\sqrt[4]{|x|}} e^{-\frac{\imath\pi}{4}}e^{-\frac{2\imath}{3}|x|^{3/2}} \, ,
\ee
where $\omega_{\bm{k}}^2 = k^2+m^{2}_{s {\rm (eff)}}$.
At late times, $x\gg 1$, this solution can be approximated as
\be
\label{e25}
\chi_{\bm{k}} \sim \frac{1}{\sqrt{2\omega_{\bm{k}}}} e^{\frac{2}{3}x^{3/2}} \, .
\ee
The solution (\ref{e25}) is a real-valued function, so the field operator can be written as
\be
\label{e26}
\hat{\chi}_{\bm{k}} = \chi_{\bm{k}} \left(\hat{a}_{\bm{k}} +\hat{a}_{\bm{k}}^\dagger\right) \, .
\ee
It also follows from Eq.~(\ref{e25}) that
\be
\label{e27}
\dot{\chi}_{\bm{k}} \sim \mu\sqrt{x} \chi_{\bm{k}} \, .
\ee
This leads to
\be
P_\chi(0,\tau) \equiv \chi_{\bm{k}=0}^2 \sim \frac{1}{2\mu\sqrt{\tau}} e^{\frac{4}{3}\tau^{3/2}}
\ee
and
\be
\label{eq:pchi2}
P_\chi(k,\tau) = \chi_{\bm{k}}^2 = P_\chi(0,\tau)e^{-k^2/k_\ast^2} \, ,
\ee
where
\be
\label{eq:kstar}
k_\ast^2 = \frac{\mu^2}{2\sqrt{\tau}} \, .
\ee
Note that the above expressions follow from the ``late-time'' behaviour \eqref{e25} and are valid for $N-N_{{\rm c}} \gg \DeltaN^{1/3}$, and hence with $\DeltaN \ll 1$, it is consistent to consider this regime while neglecting the expansion of the universe.

Combining \Eqs{eq:kstar} and~\eqref{eq:ilawa2} with $\tau=\mu t$, one finds
\be
\label{eq:kstar2}
\frac{k_\ast^2}{H_{{\rm c}}^2} =\frac12 \left(\frac{\mu}{H_{{\rm c}}} \right)^{3/2} \frac{1}{\sqrt{N-N_{{\rm c}}}}=\frac12 \frac{|m_{s {\rm (eff)}}|}{H_\mathrm{c}} \frac{1}{N-N_\mathrm{c}} \,.
\ee
Hence, the dimensionless power spectrum ${\cal P}_\chi(k,\tau) \equiv k^3/(2 \pi^2) P_\chi(k,\tau)$ assumes maximal values at scales $k\sim k_\ast$, where $k_\ast^2/H_{{\rm c}}^2$ diminishes as the instability proceeds, from values $\sim  \left(\frac{\mu}{H_{{\rm c}}} \right)^{2}$ when $N-N_{{\rm c}} \sim \DeltaN^{1/3}$ to values $\gtrsim \left(\frac{\mu}{H_{{\rm c}}} \right)^{3/2}$. At all times during the geometrical destabilisation the power spectrum thus peaks at sub-Hubble wavenumbers. 

The two-point correlator of $\chi$ and its time derivative can be calculated from \Eqs{e27} and \eqref{eq:pchi2} according to
\be
\label{eq:416}
\langle \chi^2 \rangle = \frac{1}{(2\pi)^3} \int \mathrm{d}^3\bm{k} P_\chi(k,\tau) = \frac{1}{8\pi^{3/2}} P_\chi(0,\tau) k_\ast^3 \sim \frac{1}{2^{11/2}\pi^{3/2}} \frac{\mu^2}{\tau^{5/4}} e^{\frac{4}{3}\tau^{3/2}}
\ee
and
\be
\langle \dot \chi^2 \rangle  \sim \mu^2 \tau \langle \chi^2 \rangle \sim |m^{2}_{s {\rm (eff)}}| \langle \chi^2 \rangle   \, .
\label{chi-velocity}
\ee

The exponentially growing inhomogeneities of the field $\chi$ affect the curvature perturbation, particularly at scales $k\sim k_\ast$. The contribution $\mathcal{P}_\zeta^{(\chi)}$ to the power spectrum of the curvature perturbations originating from the perturbations of the field $\chi$  can be estimated, {\it e.g.} with the use of so called $\delta N$ formalism. This computation has been performed in \Refs{Lyth:2010zq,Lyth:2012yp}, where it was found that 
\be
\label{est:lyth}
\mathcal{P}_\zeta^{(\chi)} = \frac{1}{\sqrt{\pi}} \left( \frac{H}{2m_\mathrm{max}} \right)^2 \left( \frac{\langle \dot\chi^2 \rangle}{\langle \dot\chi^2 \rangle+\dot{\phi}^2} \right)^2 \left( \frac{k}{k_\ast}\right)^3
\ee
for  $k<k_\ast$ and $\mathcal{P}_\zeta^{(\chi)}$ is negligible for $k\gg k_\ast$ because of the exponential suppression in \Eq{eq:pchi2}. The above expression relies on the validity of the relation (\ref{eq:ilawa2}), which is not satisfied
for modes leaving the Hubble radius long before the onset of geometrical destabilisation, so it applies to sufficiently large $k$ only.
In \Eq{est:lyth} all time-dependent quantities, including $k_\ast$, are evaluated at the time $t_{{\rm end}}$ that signals the end of geometrical destabilisation, when the exponential growth of $\chi$ stops or nonlinear
effects become important, whichever comes first. 
We can study this condition, expanding the equation of motion for the inflaton field (\ref{eq:phi_eom}) in powers of $\chi/M$:
\be
\label{eq:sroda1}
\ddot{\phi}+3H\dot{\phi}+V_\phi = \frac{2 \left\langle \chi^2 \right\rangle  }{M^2}V_\phi -4\frac{\left\langle \chi\dot{\chi} \right\rangle}{M^2}\dot{\phi} + \cdots \, .
\ee
Since in \Eq{eq:sroda1}, $\phi=\phi(t)$ is homogeneous, we need to average the inhomogeneous field $\chi^2$ over the observable Universe. 
As long as the right-hand side of this equation is negligible, the motion of the inflaton is not affected by geometrical destabilisation.
When the right-hand side becomes comparable to terms on the left-hand side of \Eq{eq:sroda1}, the field $\chi$ backreacts on the inflaton, which leads to slowing down the inflaton
and to the termination of geometrical destabilisation. In \Sec{sec:kin} we already discovered this effect for the component of $\bar{\chi}$ which was approximately homogeneous in
the observable universe and called it kinematical backreaction. Eq.~(\ref{eq:sroda1}) shows that the inhomogeneities in $\chi$ can also terminate geometrical destabilisation. 
Making use of \Eq{chi-velocity}, one can check that the typical amplitude of the second term in the right-hand side of \Eq{eq:sroda1} is $|m_{s {\rm (eff)}}|/H_{\mathrm{c}}$ larger than the first one, hence it provides the main source of backreaction. This implies that the inhomogeneities in $\chi$ terminate geometrical destabilisation when $\langle \chi^2 \rangle/ M^2 \sim H_{\mathrm{c}}/ |m_{s {\rm (eff)}}|$. Since $\bar{\chi}^2 \ll \langle \chi^2 \rangle$, this happens much before $\bar{\chi}^2 \sim M^2$, hence geometrical destabilisation may end because of perturbative backreaction rather than kinematical backreaction.

Given the short duration of geometrical destabilisation one can safely replace $H$ by $H_{{\rm c}}$ in \Eq{est:lyth}. As the second term in parenthesis is smaller than one, one obtains the following bound:
\be
\label{est:lythkt}
\mathcal{P}_\zeta^{(\chi)} \leq \frac{1}{\sqrt{\pi}} \left( \frac{H_\mathrm{c}}{2m_\mathrm{max}} \right)^2 \left( \frac{k}{k_\ast}\right)^3\qquad\textrm{for}\,\,k<k_\ast\,,\,\,\textrm{otherwise negligible}\,.
\ee
As $m_\mathrm{max}^2 \gg H_{{\rm c}}^2$, the bound (\ref{est:lythkt}) demonstrates that it is impossible that the curvature perturbation enters the non-linear regime with $\mathcal{P}_\zeta\sim 1$, and we therefore conclude that geometrical destabilisation appears incapable of ending inflation due to nonlinear effects.

Notice that although the amount of nonlinearities required to terminate inflation is difficult to determine, the fact that the curvature power spectrum remains smaller than one is a conservative argument for reaching the conclusion that inflation does not end. Indeed, since the power spectrum peaks at scales around $k_*$ that are much smaller than the Hubble radius, the so-called ``effective-density approximation'' of \Refs{Goldwirth:1989pr,Goldwirth:1989vz} suggests that the only effect of inhomogeneities on such scales is simply to add a contribution to the total energy density that redshifts away as $a^{-4}$, hence does not jeopardise inflation. This has been recently checked using full General Relativity codes in Ref.~\cite{East:2015ggf} (in the context of single-field inflation with a plateau-like potential, which is different from ours, but we do not expect substantial qualitative differences).

\subsection{Observable constraints on a sidetracked phase of minimal duration}

If geometrical destabilisation does not end inflation, inflation proceeds further, but along another path in field space. This sidetracked inflationary phase has been studied in \Ref{GD2}, in models in which it lasts more than about 60 $e$-folds, so that cosmological observations are unaffected by the preceding geometrical destabilisation. It leads to interesting observable signatures that can be used to constrain such phases, such as substantial primordial non-Gaussianities.

Let us discuss the case where the duration of the sidetracked phase is roughly 60 $e$-folds, in such a way that the relation (\ref{eq:ilawa2}) describing the time dependence of the effective mass $m^{2}_{s {\rm (eff)}}$ applies to the moment when the scales observable in the CMB leave out the Hubble radius, $\Delta N$ $e$-folds before the onset of geometrical destabilisation.

With $k_\ast$ given in \Eq{eq:kstar2}, one finds the contribution to the curvature power spectrum \eqref{est:lyth} for the wavenumbers $k_{{\rm CMB}}=e^{-\Delta N} H_{{\rm c}}$ corresponding to CMB scales
\be
\label{est:lythkt2}
\mathcal{P}_\zeta^{(\chi)}= \frac{1}{\sqrt{2 \pi}} \left( \frac{H_\mathrm{c}}{m_\mathrm{max}} \right)^{7/2}  e^{-3\Delta N} \Ndes^{3/2}\left( \frac{y}{1+y} \right)^2,
\ee
where $y=\langle \dot\chi^2 \rangle_{{\rm end}}/ \dot \phi^2_{{\rm end}}$ and $\Ndes=N_{{\rm end}}-N_{{\rm c}}$ is the duration of the destabilisation phase. Because of its strong scale dependence, corresponding to $n_{{\rm s}}=4$, this contribution to the curvature perturbation has to be subdominant compared to the one generated by fluctuations of the inflaton field. Simply writing $\mathcal{P}_\zeta^{(\chi)}(k_{{\rm pivot}})< \kappa A_s$, where $A_s=2.1 \times 10^{-9}$ is the observed amplitude of the almost scale-invariant curvature power spectrum, and $\kappa$ denotes a small number, one obtains the bound
\be
\Delta N > 6.4-\frac13 \ln \kappa-\frac{7}{6}\ln\left(\frac{m_\mathrm{max}}{H_\mathrm{c}}\right)+\frac12 \ln \left(\Ndes \right)+\frac23 \ln \left(\frac{y}{1+y} \right)\,.
\label{lower-bound-delta-N}
\ee
A precise estimate of the last three terms in the right-hand side of \eqref{lower-bound-delta-N} requires a detailed understanding of the end of geometrical destabilisation that is beyond the scope of this paper. For instance, with \eqref{chi-velocity} and $\dot \phi^2_{{\rm end}} \sim \dot \phi^2_{{\rm c}}=2\epsilon_{1,\mathrm{c}}H^2_\mathrm{c}\M^2$, we obtain the back-of-the-envelope estimate
\be
y \sim  \frac{m_\mathrm{max}^2}{m^2} \frac{\langle \chi^2 \rangle_{\rm end}}{M^2}\,,
\ee
where we have seen in \Sec{sec:kin} that $\langle \chi^2 \rangle_{\rm end}/M^2$ depends on the precise form of the completion of the kinetic term away from $\chi=0$,\footnote{In \Sec{sec:kin} we have seen this for the homogenous component $\chi_{{\rm end}}$ but we naturally expect a similar model-dependence for the average $\langle \chi^2 \rangle_{\rm end}$ of the inhomogeneous field.} and that even in the minimal realisation, the precise value of $m_\mathrm{max}^2/m^2$ is quite sensitive to model parameters.
Despite this, \Eq{lower-bound-delta-N} shows that although the contribution \eqref{est:lyth} to the curvature power spectrum is safely negligible compared to unity, so that the instability does not jeopardise inflation itself, its amplitude can easily be comparable to the one observed on CMB scales. Making sure that this blue-tilted component is negligible compared to $A_s$ necessitates to push $\Delta N \gtrsim {\cal O}(5)$, which makes it difficult to comply with the assumption that \Eq{eq:ilawa2} also holds at that moment. Although our discussion is qualitative, it seems to indicate that compatibility with observations severely constrains this possibility.
 
\section{Summary}
\label{sec:sum}

In this paper, we studied effects of kinematical backreaction in the simplest model of geometrical destabilisation and performed linear analysis of the perturbations
of the unstable field. We found that the simplest and most natural way in which geometrical destabilisation can be implemented seems to be in conflict with the expectation that geometrical
destabilisation actually terminates inflation. In particular, we determined that the instability is effectively shut off kinematically soon enough that the perturbations do not reach the fully non-linear regime. 
This situation is markedly different from models of reheating assisted by geometrical destabilisation \cite{KTW}, in which the instability occurs after slow-roll inflation.
While it remains to be seen with the use of more sophisticated methods, such as lattice simulations, if 
there are situations in which this conclusion can be avoided, 
our calculations point towards sidetracked inflation \cite{GD2}, where the two-field system switches to a different inflating trajectory, as a more natural fate of geometrical destabilisation. 

Let us finally note that even though the growth of field fluctuations remains in the perturbative regime during geometrical destabilisation, the bump in the power spectrum it produces at small scales might give rise to substantial amounts of primordial black holes, which could be another way to constraint the scenario of geometrical destabilisation.

\subsubsection*{Acknowledgements}

We thanks M.~Cicoli and F.~Pedro for inspiring and motivating discussions. S.RP is supported by the European Research Council (ERC) under the European Union's Horizon 2020 research and innovation programme (grant agreement No 758792, GEODESI project). K.T.~is supported by grant No.\ 2014/14/E/ST9/00152 from the National Science Centre (Poland). V.V.~acknowledges funding from the European Union's Horizon 2020 research and innovation programme under the Marie Sk\l odowska-Curie grant agreement N${}^0$ 750491.

\appendix

\section{Initial conditions} \label{sec:app1}

In this appendix, we show how to perform various analytical estimates of the spectator field variance at the onset of geometrical destabilisation: 
\begin{align}
\left\langle \chi^2_\mathrm{c} \right\rangle =  \int_{N_\ell}^{N_\mathrm{c}} \mathrm{d} N_1 \frac{H^2(N_1)}{4\pi^2} \exp\left\lbrace - \frac{2}{3}\int_{N_1}^{N_\mathrm{c}}\mathrm{d}N_2 \left[ \frac{m^2}{H^2\left(N_2\right)}-4\left(\frac{M_{{}_\mathrm{Pl}}}{M}\right)^2 \epsilon_1\left(N_2\right)\right]\right\rbrace
,\label{chi2-appendix}
\end{align}
see \Eq{chi2}.
We first make use of the slow-roll approximation, and identify the regions of models and parameter space in which it can fail. For this reason, we then perform the exact computation of \eqref{chi2-appendix} for monomial potentials, and we compare the two types of estimates.

\paragraph{Slow-roll approximation.} 
In the slow-roll approximation, we compute the integrals in Eq.~\eqref{chi2-appendix} under the assumption that the relative variations of $H$ and $\epsilon_1$ between $N_\ell$ and $N_\mathrm{c}$ are small. Before this though, one should check the self-consistency of this approach. Using \Eq{eq:ms2_2}, one obtains the exact relation 
$\epsilon_{1,\ell} H_\ell^2=\epsilon_{1,\mathrm{c}} H_{\mathrm{c}}^2 \left[ 1 - \left(\frac{H_\ell}{m}\right)^2 \right]$. Plugging into this the first-order Taylor expansion around the critical time of $\epsilon_{1,\ell} \simeq \epsilon_{1,\mathrm{c}}+\epsilon_{1,\mathrm{c}} \epsilon_{2,\mathrm{c}}(N_\ell-N_\mathrm{c})$ and $H_{\ell} \simeq H_{\mathrm{c}}-H_{\mathrm{c}} \epsilon_{1,\mathrm{c}}(N_\ell-N_\mathrm{c})$, one finds the number of e-folds elapsed in the light but stabilised phase
\be
\DeltaN \equiv N_\mathrm{c}-N_\ell  \simeq  \left(\frac{H_\mathrm{c}}{m}\right)^2 \frac{1}{\epsilon_{2,\mathrm{c}}-2 \epsilon_{1,\mathrm{c}}}\,,
\label{duration}
\ee
where one should recall that $\epsilon_{2,\mathrm{c}}>2\epsilon_{1,\mathrm{c}}$ is a prerequisite for the geometrical destabilisation to happen, and hence the relative variations
\be
\frac{\epsilon_{1,\mathrm{c}}-\epsilon_{1,\ell}}{\epsilon_{1,\mathrm{c}}} = \left(\frac{H_\mathrm{c}}{m}\right)^2 \frac{1}{1-\frac{2 \epsilon_{1,\mathrm{c}}}{\epsilon_{2,\mathrm{c}}}}\, 
\label{variation-epsilon-2}
\ee
and
\be
\frac{H_\ell-H_{\mathrm{c}}}{H_{\mathrm{c}}} = \frac{1}{2} \left(\frac{H_\mathrm{c}}{m}\right)^2 \frac{1}{\frac{\epsilon_{2,\mathrm{c}}}{   2 \epsilon_{1,\mathrm{c}}}-1}\, .
\label{variation-H-2}
\ee
Self consistency of the approach thus requires that the right hand sides of Eqs.~\eqref{variation-epsilon-2} and \eqref{variation-H-2} be small. The common factor $\left(\frac{H_\mathrm{c}}{m}\right)^2$ is indeed much smaller the unity in the models that we consider. The only case in which the slow-roll approximation is not self-consistent is hence if $\epsilon_{2,\mathrm{c}}/(2\epsilon_{1,\mathrm{c}})$ is too close to one, in the sense that $\vert \epsilon_{2,\mathrm{c}}/(2\epsilon_{1,\mathrm{c}})-1\vert \lesssim{\cal O}(H_\mathrm{c}/m)^2$. For monomial potentials $V(\phi) \propto \phi^p$, which are amenable to exact computations, we will see indeed that the slow-roll estimate becomes inaccurate for $p$ very close to $2$, corresponding to the limiting case $\epsilon_{2}=2\epsilon_{1}$. Having delineated the regime of validity of the slow-roll approximation, it is straightforward to use the Taylor expansions given above to perform the integrals in \Eq{chi2-appendix}, finding
\be
\left\langle \chi^2_\mathrm{c} \right\rangle \simeq  \left( \frac{H_\mathrm{c}}{2\pi} \right)^2 \left\lbrace \frac{1}{2}\sqrt{3 \pi \DeltaN}\mathrm{erf}\!\left[\sqrt{\DeltaN/3} \right]-3 \epsilon_{1,\mathrm{c}} \DeltaN \left[ e^{-\DeltaN/3}-1\right] \right\rbrace ,
\label{eq:ktd2}
\ee
where $\DeltaN$ is given in Eq.~\eqref{duration}. \Eq{eq:ktd2}, and the limiting behaviours (\ref{eq:ktd3}), are general and can be applied to any inflationary model under the self-consistency condition expressed above. On the other hand, specifying a model can provide a relation between $\epsilon_{1,\mathrm{c}}$ and $\epsilon_{2,\mathrm{c}}$,
which in turn allows for expressing $\langle \chi^2_\mathrm{c} \rangle$ in terms of the parameters describing the geometrical destabilisation. For example, in Starobinsky inflation, 
\be
V(\phi)=\frac{3}{4}\Lambda^2M_P^2\left[1-\mathrm{exp}\left(-\sqrt{\frac{2}{3}}\frac{\phi}{\M}\right)\right]^2 \, ,
\label{eq:Starobinsky}
\ee
we have $\epsilon_{2,\mathrm{c}}^2 = \frac{16}{3}\epsilon_{1,\mathrm{c}} \ll 1$, in which case Eqs.~\eqref{eq:epsilonc} and \eqref{duration} give $\DeltaN= \frac{\sqrt{3}}{2} \left( \frac{H_\mathrm{c}}{m}\right)^3  \frac{\M}{M}$, and hence
\be
\left\langle \chi_{\mathrm{c}}^2 \right\rangle_{\mathrm{Starobinsky}}\simeq
\left\{
\begin{array}{ll}
\left( \dfrac{H_\mathrm{c}}{2\pi} \right)^2 \dfrac{\sqrt{3}}{2} \left( \dfrac{H_\mathrm{c}}{m}\right)^3  \dfrac{\M}{M} \quad \textrm{for}\ \left(\dfrac{H_{\mathrm{c}}}{m}\right)^{3}\dfrac{\M}{M}  \ll 1 \\
\left( \dfrac{H_\mathrm{c}}{2\pi} \right)^2 \dfrac{3^{3/4}\pi^{1/2}}{2^{3/2}} \left( \dfrac{H_\mathrm{c}}{m}\right)^{3/2} \left( \dfrac{\M}{M} \right)^{1/2} \quad \textrm{for}\ \left(\dfrac{H_{\mathrm{c}}}{m}\right)^{3}\dfrac{\M}{M}  \gg 1 
\end{array}
\right.
\, .
\ee

\paragraph{Monomial potentials.} In this section, we consider the specific class of monomial potentials 
\be
\label{eq:pot:lfi}
V\left(\phi\right)=M^4\left(\frac{\phi}{\M}\right)^p\, ,
\ee
and perform the integrals in Eq.~\eqref{chi2-appendix} analytically. We consider the range of parameters $0<p< 2$ for which the geometrical destabilisation takes place, as discussed in \Sec{sec:minimal}. At leading order in slow roll, one has
\be
\label{eq:H(N):LFIp}
H(N)=H_{\mathrm{end}} \left[1+\frac{4}{p}\left(N_{\mathrm{end}}-N\right)\right]^{p/4},
\ee
and hence
\be
\label{eq:epsilon(N):LFIp}
\epsilon_1(N) = \left(\frac{H_{\mathrm{end}}}{H}\right)^{4/p}\,\quad {\rm and} \quad \epsilon_2(N)=\frac{4}{p} \epsilon_1(N)\,,
\ee
where $H_{\mathrm{end}}$ denotes the value of $H$ at the end of inflation defined by $\epsilon_1=1$, \textit{i.e.} in the absence of geometrical destabilisation (of course, we will only apply these equations for $N<N_\mathrm{c}$). \\

Let us now calculate $H_{\ell}$. In monomial potentials, $m^{2}_{s {\rm (eff)}}/H^2$ formally always go to zero in the infinite past, so the field always starts out being light. In some cases (that we will determine now), it then becomes heavy, then it becomes light again before being destabilised. 
From \Eqs{eq:ms2_2}, \eqref{eq:epsilonc} and \eqref{eq:H(N):LFIp}, requiring that $m^{2}_{s {\rm (eff)}} = H_\ell^2$ leads to
\be
\left(\frac{H_\ell}{H_{\mathrm{c}}}\right)^{2-\frac{4}{p}} = 1-\left(\frac{H_\ell}{m}\right)^2\, .
\label{equ:Hl}
\ee
This equation has, in general, no analytical solution for $H_\ell$. In the limit where $m \gg H_{\ell}$ however it can be solved perturbatively and one obtains, at leading order in $H_{\mathrm{c}}/m$,
\be
\label{eq:LFIp:Hl:pert}
H_\ell \simeq H_{\mathrm{c}}\left[1+\frac{\left(\frac{H_{\mathrm{c}}}{m}\right)^2}{\frac{4}{p}-2}\right]\, .
\ee
Obviously, this result holds for $H_{\mathrm{c}}/m \ll \sqrt{2-p}$ (see below for when this is not true). Notice that this is consistent with Eq.~(\ref{variation-H-2}). For example, if $p=1$, the equation can be solved exactly and one finds
\be
\left.H_{\ell} \right\vert_{p=1}= \frac{m}{\sqrt{2}}\sqrt{1-\sqrt{1-4\left(\frac{H_{\mathrm{c}}}{m}\right)^2}}.
\ee
If one expands this expression in $H_{\mathrm{c}}/m\ll 1$, one finds
\be
\left.H_{\ell} \right\vert_{p=1}= H_{\mathrm{c}}\left[1+\frac{1}{2}\left(\frac{H_{\mathrm{c}}}{ m^2} \right)^2+\frac{7}{8}\left(\frac{H_{\mathrm{c}}}{ m^2} \right)^4+\cdots\right]
\ee
which is indeed consistent with the above expression. In the case where $p$ is so close to $2$ that $H_{\mathrm{c}}/m \gg \sqrt{2-p}$, \Eq{equ:Hl} can be solved by performing an expansion in $2-p$. This gives rise to $(p-2)\ln(H_\ell/H_{\mathrm{c}}) + H_\ell^2/m^2\simeq 0$. After a few manipulations this can be solved with the 0-branch of the Lambert function and one obtains
\be
H_\ell \simeq m \sqrt{\frac{p-2}{2}W_0\left[\frac{2}{p-2}\left(\frac{H_\mathrm{c}}{m}\right)^2\right]}\, .
\label{W0}
\ee
Recalling that $W_0(x)\simeq x$ when $x\ll 1$, one finds that if 
$(H_\mathrm{c}/m)^2\ll 2-p$, in which case \Eq{W0} is a priori not applicable, it does give
$H_\ell \simeq H_{\mathrm{c}}$, which is consistent with the previous regime. In the opposite limit however, the Lambert function is not defined as soon as its argument is less than $-1/e$. In that case, since the field is always light (since it is massless) at the onset of GD, the fact that the equation has no solution implies that it never transits from heavy to light, in other words, that it remains light throughout the entire evolution. This happens when
\be
2-p< 2e \left(\frac{H_\mathrm{c}}{m}\right)^2. 
\ee 
In that case, $H_\ell$ can be taken at the onset of inflation, $H_\ell = H_{\mathrm{start}}$. Note that this case is rather special though. In particular, it arises because the mass parameter $m$ is taken to be constant in the simple Lagrangian \eqref{minimal} while $H$ grows unboundedly in the past. One could expect, maybe more realistically, that such a mass parameter receives loop corrections proportional to $H$ for instance, that would change the situation. It is nonetheless interesting to consider this limiting region of parameter space and draw its full consequences.

Let us now work out the integrals in \Eq{chi2-appendix}. First we perform the change of integration variable $N\rightarrow x \equiv H/H_{\mathrm{c}}$, giving
\begin{align}
\left\langle \chi^2_{\mathrm{c}} \right\rangle & = 
 \frac{H^2_{\mathrm{c}}}{4\pi^2 \epsilon_{1,\mathrm{c}}}
  \int^{H_\ell/H_{\mathrm{c}}}_{1} 
   \mathrm{d} x_1 x_1^{4/p+1}  \exp\left\lbrace
 -\frac{2 m^2}{3 H^2_{\mathrm{c}}\epsilon_{1,\mathrm{c}}}  \int^{x_1}_{1}\mathrm{d}x_2 x_2^{4/p-3}
+\frac{8}{3} \left(\frac{M_{{}_\mathrm{Pl}}}{M}\right)^2 \ln\left( x_1 \right)
\right\rbrace \\
&= \frac{H^2_{\mathrm{c}}}{4\pi^2 \epsilon_{1,\mathrm{c}}}
 \exp\left[
 \frac{m^2}{3 H^2_{\mathrm{c}}\epsilon_{1,\mathrm{c}}\left(\frac{2}{p}-1\right)} 
\right]
  \int^{H_\ell/H_{\mathrm{c}}}_{1} 
   \mathrm{d} x_1 x_1^{\frac{4}{p}+1+\frac{8}{3} \left(\frac{M_{{}_\mathrm{Pl}}}{M}\right)^2}  \exp\left[-
 \frac{m^2 x_1^{\frac{4}{p}-2}
}{3 H^2_{\mathrm{c}}\epsilon_{1,\mathrm{c}}\left(\frac{2}{p}-1\right)} \right]
 \, .
\end{align}
Using the further change of integration variable $x_1 \to y(x_1) \equiv \frac{m^2 x_1^{\frac{4}{p}-2}
}{3 H^2_{\mathrm{c}}\epsilon_{1,\mathrm{c}}\left(\frac{2}{p}-1\right)} $, 
the resulting integral can be expressed in terms of the incomplete gamma function, and one obtains
\begin{align}
\left\langle \chi^2_{\mathrm{c}} \right\rangle & = 
 \frac{H^2_{\mathrm{c}}}{4\pi^2 \epsilon_{1,\mathrm{c}}}
e^{y(1)}
 y(1)^{-s} \frac{1}{4/p-2}
 \Gamma\left[s,y\left(1\right),y\left(\frac{H_\ell}{H_{\mathrm{c}}}\right)\right]
 \, ,
 \label{dispersion-monomials}
\end{align}
where we have defined
\be
s \equiv \frac{1+\frac{2}{p}+\frac{4}{3}\left(\frac{M_{\mathrm{Pl}}}{M}\right)^2}{\frac{2}{p}-1}\,.
\ee
We have 
\be
\begin{cases}
\frac{s}{y(1)}=1+\frac{3}{4}\left(1+\frac{2}{p}\right)\left(\frac{M}{\M}\right)^2\\
\frac{s}{y\left(\frac{H_\ell}{H_{\mathrm{c}}}\right)}=\left[1+\frac{3}{4}\left(1+\frac{2}{p}\right)\left(\frac{M}{\M}\right)^2\right] \left(\frac{H_\ell}{H_{\mathrm{c}}}\right)^{2-4/p}
\end{cases},
\ee
so that, except for the highly particular case of $H_{\mathrm{start}}/H_{\mathrm{c}} \gg e^{p/[2(2-p)]}$, one can write $y(1)=s(1+\alpha_1)$ and $y\left(\frac{H_\ell}{H_{\mathrm{c}}}\right)=s(1+\alpha_\ell)$ with $\alpha_1$ and $\alpha_\ell$ much smaller than unity. One could then think of using the expansion $\Gamma[s,s(1+\alpha_1),s(1+\alpha_\ell)]\simeq -e^{-s} s^s (\alpha_1-\alpha_\ell)$. However, we have checked numerically that for the typical parameters involved, any finite-order Taylor expansion is inaccurate, so that we could not find any further analytical insight beyond \eqref{dispersion-monomials}.

\paragraph{Comparison.} Let us now compare the expression \eqref{dispersion-monomials} for monomial potentials with the generic ones obtained in \Eq{eq:ktd3} under the slow-roll approximation, and see when the later provide a good approximation to the former. 
Because of the second \Eq{eq:epsilon(N):LFIp}, the generic expression \eqref{duration} for the duration of the light stabilised phase gives
\begin{align}
\DeltaN & \simeq  \frac{2p}{2-p}\left(\frac{H_{\mathrm{c}}}{m}\right)^4\left(\frac{\M}{M}\right)^2\,.
\label{DeltaNmonomial}
\end{align}
One can check that the expression \eqref{DeltaNmonomial} can also be derived from the relations~(\ref{eq:H(N):LFIp})-\eqref{eq:epsilon(N):LFIp} and \eqref{eq:LFIp:Hl:pert} that are specific to monomial potentials. 
Using \Eq{DeltaNmonomial}, \Eq{eq:ktd3} then reads
\begin{align}
\left\langle \chi^2_{\mathrm{c}} \right\rangle_{\mathrm{according\ to\ Eq.~(\ref{eq:ktd3})}} = 
\begin{cases}
\dfrac{p}{4}\dfrac{H_{\mathrm{c}}^2}{\pi^2}\left(\dfrac{H_{\mathrm{c}}}{m}\right)^4\left(\dfrac{\M}{M}\right)^2
\quad \text{if}\ \sqrt{2-p}\gg \dfrac{\M}{M}\left(\dfrac{H_{\mathrm{c}}}{m}\right)^2\\
\dfrac{H_{\mathrm{c}}^2}{4 \pi^2} \sqrt{\dfrac{3\pi}{2}\dfrac{p}{2-p}}\left(\dfrac{H_{\mathrm{c}}}{m}\right)^2\dfrac{\M}{M}
\quad \text{if}\ \dfrac{M}{\M}\ll \sqrt{2-p}\ll \dfrac{\M}{M}\left(\dfrac{H_{\mathrm{c}}}{m}\right)^2\ ,\\
\dfrac{H_{\mathrm{c}}^2}{4 \pi^2} \dfrac{3p}{2(2-p)} \left(\dfrac{H_{\mathrm{c}}}{m}\right)^2
\quad \text{if}\ \dfrac{M}{\M}\gg \sqrt{2-p}
\end{cases}
\label{dispersion-monomial-slow-roll}
\end{align}
where we note that the two limiting values indeed verify $M/\M \ll (\M/M) (H_{\mathrm{c}}/m)^2$ by virtue of \Eq{eq:epsilonc}. Note however that $(\M/M) (H_{\mathrm{c}}/m)^2$ can be larger than unity in general, in which case $p$ can not fulfil the condition on the first line of \eqref{dispersion-monomial-slow-roll}. 

\begin{figure}
\begin{center}
\includegraphics*[width=7.1cm]{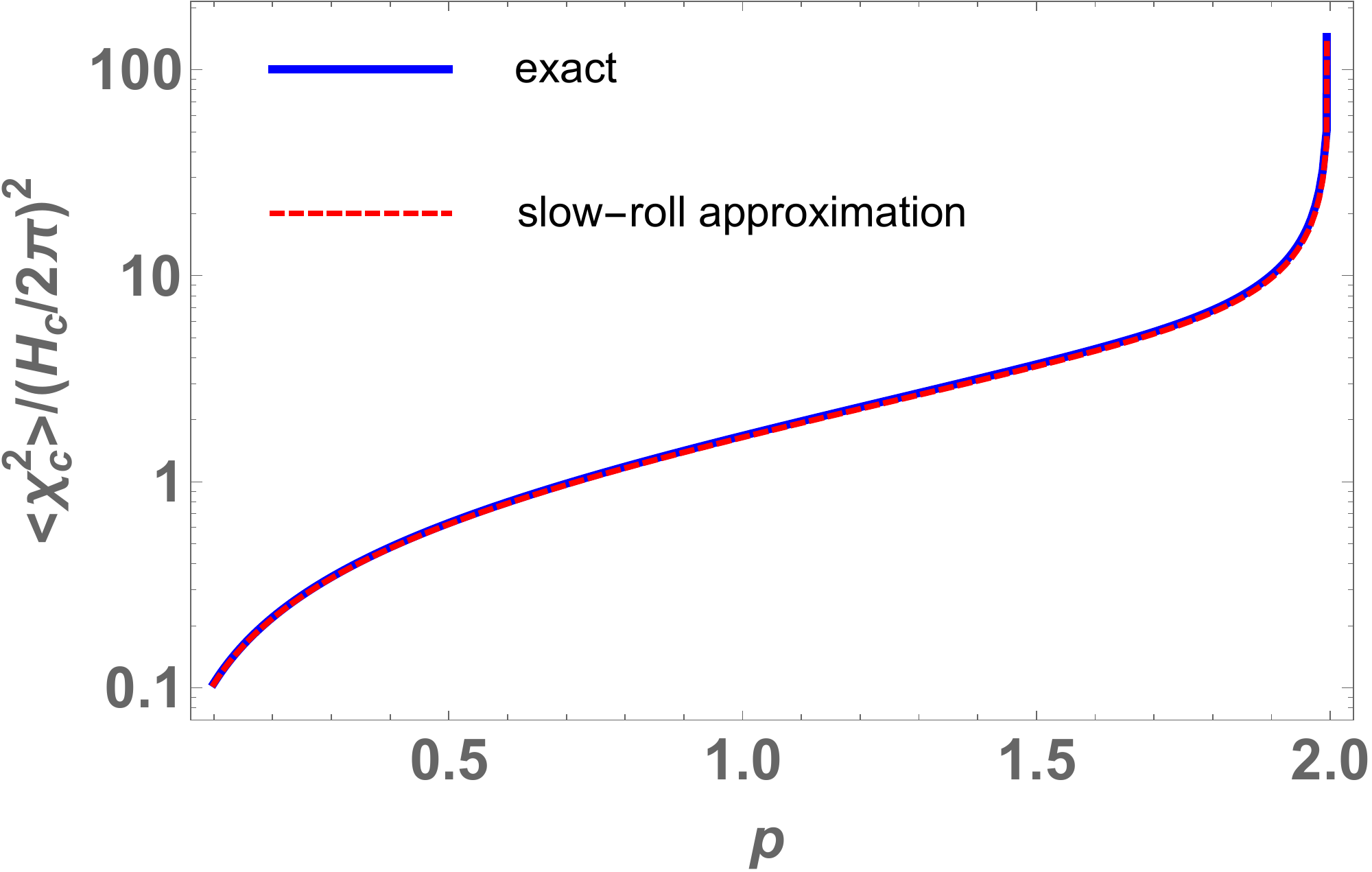}
\includegraphics*[width=7.2cm]{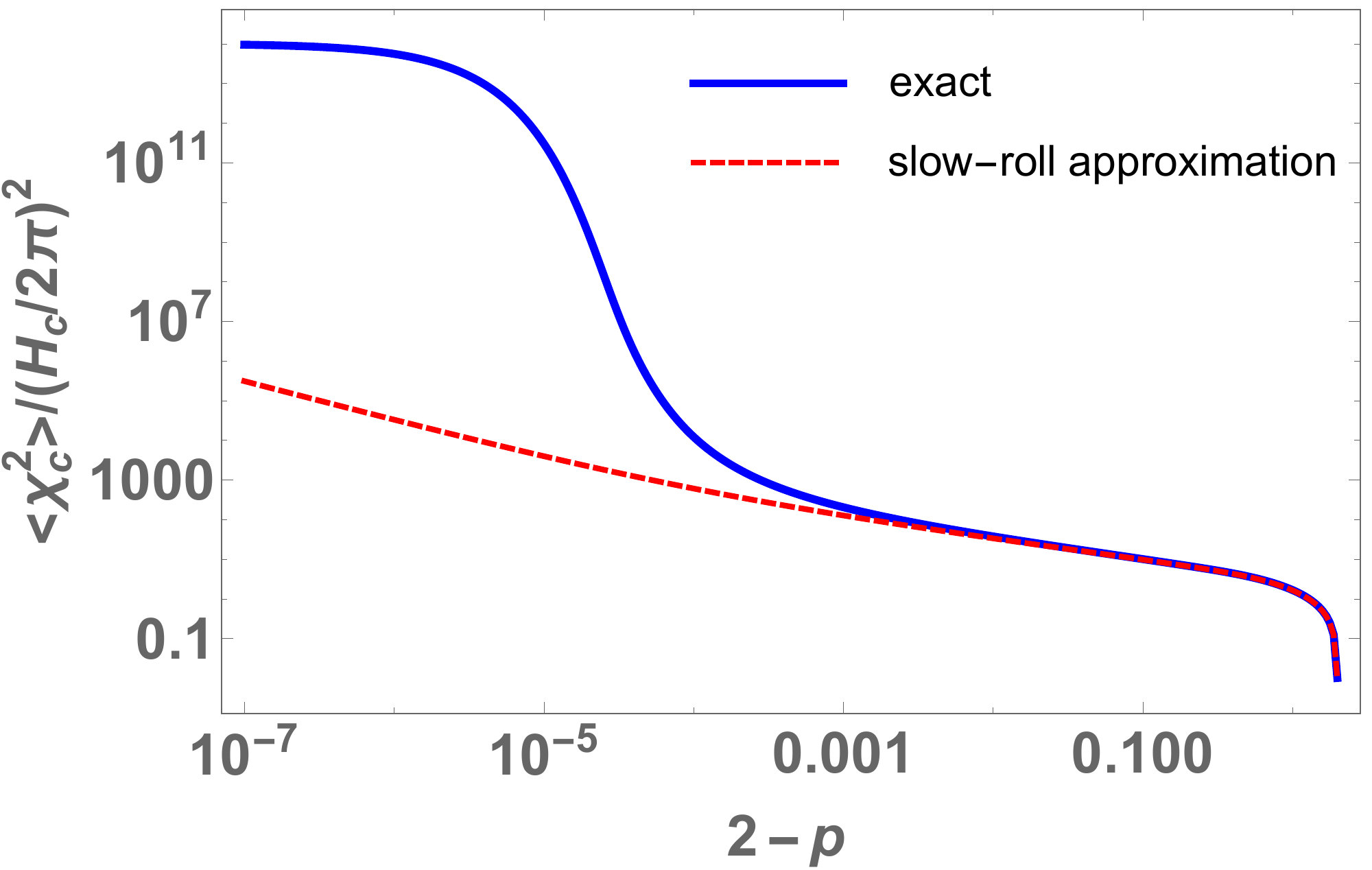}
\caption{Spectator field variance $\langle\chi_\mathrm{c}^2\rangle$ at the beginning of geometrical stabilisation for the monomial inflationary potential~(\ref{eq:pot:lfi}) with $M= 10^{-2} \M$, $m=10 H_{\mathrm{c}}$ and $H_{\mathrm{start}}=10^3 H_{\mathrm{c}}$. We compare the slow-roll result \eqref{eq:ktd1} and the exact result \eqref{dispersion-monomials} as a function of $p$. The agreement between the two results is excellent, except for very small values of $2-p$, close to which the right panel allows one to zoom in since it uses a logarithmic scale on $2-p$.
\label{fig:slowroll-vs-exact}}
\end{center}
\end{figure}

In \Fig{fig:slowroll-vs-exact}, we display the slow-roll result \eqref{eq:ktd1}-\eqref{dispersion-monomial-slow-roll} and the exact result \eqref{dispersion-monomials} as a function of $p$, for $M= 10^{-2} \M$, $m=10 H_{\mathrm{c}}$ and $H_{\mathrm{start}}=10^3 H_{\mathrm{c}}$. One can see that for $2-p \lesssim 10^{-4}$, the slow-roll result underestimates the dispersion. However, for larger values, the agreement between the two estimates is excellent, and this agreement actually extends beyond the expected self-consistency regime of the slow-roll approach $\frac{H_{\mathrm{c}}}{m} \ll \sqrt{2-p}$.

\end{document}